\documentclass[12pt]{article}
\usepackage{amsmath,amssymb,bm,graphicx,epsfig,subfigure,makecell}
\usepackage{lineno,booktabs,url,color}
\usepackage{graphicx}
\usepackage{subfigure}

\usepackage[numbers,sort&compress]{natbib}

\def\ltap{\raisebox{-.6ex}{\rlap{$\,\sim\,$}} \raisebox{.4ex}{$\,<\,$}}

\def\tL{{\tilde L}}

\def\beqn{\begin{eqnarray}}
\def\eeqn{\end{eqnarray}}
\def\beq{\begin{equation}}
\def\eeq{\end{equation}}

\usepackage{hyperref}
\begin{document}
\include{epsf}
\begin{titlepage}

\renewcommand{\thefootnote}{\fnsymbol{footnote}}
\vspace*{1cm}

\begin{center}
{\Large \bf
  Drell--Yan lepton pair production at low invariant masses:\!\!\\\
 transverse-momentum resummation
  \\
  and non-perturbative  effects in QCD \\
}
\end{center}

\par \vspace{2mm}
\begin{center}
  {\bf Stefano Camarda${}^{(a)}$},
  {\bf Giancarlo Ferrera${}^{(b)}$} and
  {\bf Lorenzo Rossi${}^{(b)}$}\\

\vspace{5mm}

${}^{(a)}$
CERN, CH-1211 Geneva, Switzerland\\\vspace{1mm}
${}^{(b)}$
Dipartimento di Fisica, Universit\`a di Milano
and\\
INFN, Sezione di Milano, I-20133 Milan, Italy\\\vspace{1mm}

\end{center}

\vspace{1.5cm}

\par \vspace{2mm}
\begin{center} {\large \bf Abstract} \end{center}
\begin{quote}
  \pretolerance 10000
  We consider the transverse-momentum ($q_T$) distribution of Drell-Yan lepton pairs produced with invariant masses ($M$)
  from low values up to the $Z$-boson peak ($4\leq M \leq 116$~GeV).
  We present perturbative predictions obtained by consistently combining the
  resummation of logarithmically enhanced QCD corrections at small $q_T$ ($q_T\ll M$) up to next-to-next-to-next-to-next-to-leading logarithmic accuracy
  with the available fixed-order calculations at  next-to-next-to-leading order (i.e.\ $\mathcal{O}(\alpha_S^3)$) valid at large $q_T$.
  For very low $q_T$ ($q_T\sim \Lambda_{\mathrm{QCD}}$), non-perturbative (NP) QCD effects
  become dominant and have been included through a NP form factor with a small number of free-parameters.
  We compare our results with multiple experimental datasets from hadron colliders, finding excellent agreement between theory and data.
  By fitting the  NP parameters, we achieve a precise extraction of the NP form factor and the so-called Collins-Soper kernel.

\end{quote}

\vspace*{\fill}
\vspace*{1.75cm}

\begin{flushleft}
August 2025
\end{flushleft}
\end{titlepage}

\setcounter{footnote}{1}
\renewcommand{\thefootnote}{\fnsymbol{footnote}}

\section{Introduction}

The production of Drell–Yan (DY) lepton pairs with high invariant masses ($M\gg \Lambda_{\mathrm{QCD}}$), mediated by a neutral gauge boson ($\gamma^*/Z$),
is a benchmark process measured with high precision at hadron colliders~\cite{Holmes:2011ey,Evans:2008zzb}.
It provides important tests of the Standard Model (SM) through precise measurements of its fundamental parameters~\cite{Camarda:2016twt,CMS:2011utm,ATLAS:2017rzl,CDF:2022hxs}
and also imposes constraints on new physics~\cite{CidVidal:2018eel}. Thus, accurate theoretical predictions for DY production cross sections and
corresponding kinematical distributions are essential. Among these distributions, the transverse momentum ($q_T$)
spectrum of the lepton pair plays a special role. In particular, the $q_T$ spectrum of the $Z$ boson provides key
insights into the $q_T$ spectrum of the $W$ boson, whose uncertainty, in turn, directly impacts the measurement of the $W$ boson mass~\cite{CDF:2022hxs,ATLAS:2017rzl,Rottoli:2023xdc,CDF:2013dpa,LHCb:2021bjt}.

In the region where $q_T \simeq M$, the QCD perturbative
series is controlled by a small expansion parameter, $\alpha_S(M^2)$, and the fixed-order expansion in the QCD coupling $\alpha_S$ is
theoretically justified.  In the intermediate- and low-$q_T$ region, $\Lambda_{\mathrm{QCD}} \ll q_T \ll M$,
large logarithmic corrections of infrared origin, of the form $\alpha_S^n/q_T^2\ln^m(M^2/q_T^2)$ (with $0\leq m \leq 2n-1$), spoil the convergence of the
fixed-order expansion. In this region, resummation techniques become mandatory.
Finally, at very low $q_T$, $\Lambda_{\mathrm{QCD}} \ltap q_T$, non-perturbative (NP)
effects become dominant and cannot be neglected.
However, these regions are in general not well separated in real (experimental) cases.
Therefore, to obtain reliable and accurate predictions, it is essential
to develop a theoretically consistent formalism capable of describing all the above regions in
an unified and consistent framework.

Various different formalisms have been developed with the goal of performing  all-order resummation of logarithmically-enhanced terms at small $q_T$\,\cite{Dokshitzer:1978yd,Parisi:1979se,Collins:1984kg,Catani:2010pd,Becher:2010tm,Ebert:2016gcn,Scimemi:2017etj,Bizon:2018foh,Bizon:2019zgf,Becher:2019bnm,Bacchetta:2025ara,Alioli:2021qbf,Ebert:2020dfc,Billis:2024dqq}.
In this paper we follow the general method for transverse-momentum resummation of colorless high mass systems $M\gg \Lambda_{\mathrm{QCD}}$ developed in Refs.\,\cite{Catani:2000vq,Bozzi:2005wk,Bozzi:2007pn} and applied
in the case of Higgs, vector boson and diboson production \,\cite{Catani:2015vma,Catani:2018krb,Cieri:2015rqa,Grazzini:2015wpa,deFlorian:2011xf,deFlorian:2012mx,Bozzi:2010xn}. This formalism has been implemented, in the case of Drell--Yan production,
in the {\ttfamily DYTurbo} public numerical program\,\cite{Camarda:2019zyx,WEBLINK} starting from the numerical implementations of
Refs.\,\cite{Bozzi:2010xn,Catani:2015vma}.

Phenomenological studies of the $q_T$ spectrum of lepton pairs with an invariant mass ($M$) of the order of the $Z$-boson mass
$M\sim m_Z$ at the Tevatron and the LHC, using the formalism of  Refs.\,\cite{Catani:2000vq,Bozzi:2005wk,Bozzi:2007pn}, have been performed in \cite{Catani:2015vma,Camarda:2021ict,Camarda:2023dqn}.

In particular, it has been shown  in\,\cite{Catani:2015vma}, that $q_T$ resummed (and matched) predictions within a purely perturbative approach for the $q_T$ recoil dynamics
give a good description, within the perturbative uncertainties, of the experimental data from Tevatron and LHC  for $M\sim m_Z$ except
at very low $q_T$ values ($q_T \sim 1$\,GeV).

In this paper, we extend the analysis of Ref.\,\cite{Catani:2015vma}, also considering the region of lower invariant masses $M < m_Z$.
This region of invariant masses is more delicate. On the one hand, since the QCD coupling $\alpha_S(M^2)$
increases, higher order corrections become more relevant. On the other hand, approaching the region of very low energies,
a breakdown of perturbation theory  and the dominance of
NP effects
is expected\,\cite{Catani:2015vma}.  In fact, studies of DY transverse momentum distributions at low masses with various theoretical approaches
and different perturbative accuracy showed contradictory results~\cite{Bacchetta:2019tcu,Gauld:2021pkr}.
Within the present analysis, we confirm the results from Ref.~\cite{Catani:2015vma},
observing that purely perturbative predictions are able to describe, within the errors, experimental data down
to $q_T \sim 1$ GeV. Below that region, perturbative predictivity deteriorates and a good description of the data can be obtained only by including NP effects.

NP effects at low $q_T$ can also be incorporated through the formalism of Transverse Momentum Dependent (TMD) Parton Distribution Functions (PDFs), which encode both the longitudinal and transverse momentum distributions of partons inside hadrons. This framework was developed in Refs.\cite{Collins:1989gx,Collins:1981uk,Collins:1984kg,Collins:1985ue}.
The nonperturbative components of TMDs are typically modeled using ansatz functions that depend on several NP parameters to be extracted from the experimental data. For example, the analyses in Refs.\cite{Moos:2025sal,Bacchetta:2024qre,Bacchetta:2025ara} employ several tens of free parameters to describe NP effects.

In this paper, we adopt a more minimalistic approach. We incorporate NP effects in the impact-parameter space (i.e., the Fourier conjugate of $q_T$) by introducing an NP form factor that depends on only four free parameters. Despite its simplicity, this approach provides an excellent description of the experimental data, achieving a fit quality of $\chi^2 \sim 1$ over a wide kinematic range: $4 \leq M \leq 116$~GeV and $0 \leq q_T/M \leq 0.3$.

\section{Transverse-momentum resummation and non-perturbative effects}

We briefly review the resummation formalism
developed in Refs.\,\cite{Bozzi:2005wk,Bozzi:2010xn,Catani:2015vma}
focussing on the model for NP effects.
We consider the process
\beqn
h_1 + h_2 \to Z/\gamma^* +X \to l_3+l_4+X,
\eeqn
where the $Z/\gamma^*$ boson is
produced
by the colliding hadrons $h_1$ and $h_2$, while $l_3$ and $l_4$ are the
final state leptons produced by the decay of the $Z/\gamma^*$.

The fully differential hadronic
cross section  in the lepton kinematics, $d \sigma_{h_1h_2\to l_3l_4}$
is completely specified by six independent variables.
We consider the differential cross section as a function of
 the two-dimensional transverse-momentum ${\bf q_T}$
 ($q_T=\sqrt{{\bf q_T}^2}$), the rapidity $y$
 and the invariant
mass $M$ of the lepton pair, and by two generic variables ${\bf \Omega}$
specifying the angular distribution of the leptons
with respect to the vector boson momentum.
We perform the resummation at the level of partonic cross-section.
The
differential partonic cross section is decomposed as follows:
\beqn
\label{partXS2}
    d \hat\sigma_{a_1a_2\to l_3l_4}= d \hat\sigma^{({\rm res.})}_{a_1a_2\to l_3l_4}+
    d \hat\sigma^{({\rm fin.})}_{a_1a_2\to l_3l_4}
\eeqn
where $a_1$ and $a_2$ are the partonic indices and the
first term on the right-hand side
is the resummed component
which contains (and resums) all the logarithmically-enhanced contributions of the type
$\alpha_S^n\,M^2/q_T^2\ln^m(M^2/q_T^2)$ (with $0\leq m \leq 2n-1$),
while the second term is the finite component to
be evaluated at fixed order in perturbation theory.
More precisely, the finite component is defined in
such a way that order-by-order in perturbation theory it fulfills
the following equation\,\cite{Bozzi:2005wk}:
\begin{equation}
\lim_{Q_T\to 0} \int_0^{Q_T^2}dq_T^2
\left[ \frac{d \hat\sigma^{({\rm fin.})}}{dq_T^2}
\right]=0\,.
\end{equation}
This implies that all the constant terms in $b$ space
(corresponding to $\delta(q_T^2)$ terms in $q_T$ space)
 are included,
together with the enhanced logarithms,
in $d \hat\sigma^{({\rm res.})}$.
We evaluate the finite (or remainder) term by subtracting the expansion of the resummed part from the standard fixed-order perturbative expression of the partonic cross section at the same order.

This matching procedure between resummed and fixed-order results (together with the perturbative unitarity constraint we implement on the total cross section) avoids the introduction of
large and unjustified higher-order terms at large $q_T$
and allows us to obtain, in such a region, a matched result which
is consistent with the fixed-order expansion.

The resummation is performed in the impact-parameter space $b$ (conjugated to $q_T$)\,\cite{Parisi:1979se} and the resummed
component at partonic level is written in the following factorized
form\,\,\cite{Bozzi:2005wk}\,\footnote{Resummation of large logarithms is better expressed by working in the Mellin ($N$-moment) space. For the sake of simplicity we omit
the explicit dependence on Mellin indices and on flavour indices.}:
\begin{align}
\label{resum}
{d{\hat \sigma}_{a_1a_2\to l_3l_4}^{(\rm res.)}}
&= \sum_{q}
{d{\hat \sigma}^{(0)}_{q {\bar q} \to l_3l_4}} \;
{\cal H}_{a_1a_2 \to V}\left(M,
\alpha_S;\mu_R,\mu_F,Q \right)
\times \exp\{{\cal G}(\alpha_S,\tL;\mu_R,Q)\}\,, 
\end{align}
where
\begin{equation}
\label{e:Gfactor}
   {\cal G}(\alpha_S,\tL)\ =  - \int_{b_0^2/b^2}^{Q^2}  \frac{dq^2}{q^2} \Big{[} A(\alpha_S(q^2)) \ln \frac{M^2}{q^2} + B(\alpha_S(q^2))+
   2\beta(\alpha_S(q^2))\frac{d \ln C(\alpha_S(q^2))}{d\ln\alpha_S(q^2)}
   +2\gamma(\alpha_S(q^2))\Big{]}\,,
\end{equation}
and we have introduced
the logarithmic expansion parameter $\tL\equiv \ln ({Q^2 b^2}/{b_0^2}+1)$
with $b_0=2e^{-\gamma_E}$ ($\gamma_E=0.5772...$ is the Euler number).
The scales $\mu_R$ and $\mu_F$ are the customary
renormalization and factorization scales,
while $Q\sim M$  is the resummation scale whose variations can be used
to estimate the effect of uncalculated logarithmic terms at higher orders\,\cite{Bozzi:2003jy}.

The function $A(\alpha_S)$ controls the exponentiation of the
double-logarithmic
series, while the single logarithms are controlled by the functions $B(\alpha_S)$,  $C(\alpha_S)$ and $\gamma(\alpha_S)$.
The functions  $C(\alpha_S)$ and $\gamma(\alpha_S)$ (which depend
on Mellin and flavour indices) are the collinear
functions\,\cite{Catani:2022sgr} and the moments of the customary Altarelli--Parisi splitting functions.

The functions appearing in Eq.~\eqref{e:Gfactor} are perturbative
series in $\alpha_S$:
\begin{eqnarray}
 A(\alpha_S) = \sum_{n=1}^\infty
 \Big(\frac{\alpha_S}{\pi}\Big)^n A^{(n)}\, , \qquad
 B(\alpha_S) = \sum_{n=1}^\infty \Big(\frac{\alpha_S}{\pi}\Big)^n B^{(n)}\, , \label{eqAB}\\
 C(\alpha_S) = \sum_{n=0}^\infty \Big(\frac{\alpha_S}{\pi}\Big)^n C^{(n)}\, ,  \qquad
 \gamma(\alpha_S) = \sum_{n=1}^\infty \Big(\frac{\alpha_S}{\pi}\Big)^n \gamma^{(n)}\, , \label{eqCg}
\end{eqnarray}
and $\beta(\alpha_S)$ is the QCD $\beta$ function:
\begin{equation}
    \frac{d\ln\alpha_S(\mu_R^2)}{d\ln \mu_R^2}=\beta(\alpha_S(\mu_R))\,.
\end{equation}
The factor $d{\hat \sigma}^{(0)}_{q {\bar q}\to l_3l_4}$ in Eq.\,(\ref{resum})
is the Born level differential cross section for the quark-antiquark annihilation
subprocess $q {\bar q} \to V \to l_3l_4$.
The functions ${\cal H}_{a_1a_2 \to V}$\,\cite{Catani:2000vq,Catani:2013tia,Catani:2012qa,Gehrmann:2012ze}
are (process dependent) coefficients that contain
the hard-collinear contributions
and have a standard fixed-order expansion in powers of
$\alpha_S=\alpha_S(\mu_R^2)$.

The universal (process independent) form factor $\exp\{{\cal G}\}$
contains and resums in an exponentiated form all the $\alpha_S^k\tL^{k}$ (with $k\geq 1$) logarithmic terms
that order-by-order in $\alpha_S$ are logarithmically divergent
as $b \to \infty$ (i.e.\ $q_T\to 0$).

The N$^k$LL+N$^k$LO accuracy in the small-$q_T$ region means that we include all the logarithmic terms of the type $\alpha_S^n\tL^{n+1}$ down to $\alpha_S^n\tL^{n+1-k}$ in the function ${\cal G}$ and we evaluate
the  functions ${\cal H}_{a_1a_2 \to V}$ up to N$^k$LO (i.e.\ $\mathcal{O}(\alpha_S^k)$). Following Ref.\,\cite{Camarda:2023dqn} we have reached the N$^4$LL+N$^4$LO approximated accuracy, where the missing part of the ${\cal H}_{a_1a_2 \to V}$ coefficients at $\mathcal{O}(\alpha_S^4)$ has been approximated.

We have then consistently matched the resummed results with the fixed-order calculations at high $q_T$, including finite terms in Eq.~\eqref{partXS2} up to $\mathcal{O}(\alpha_S^3)$. These contributions are now implemented in the {\ttfamily DYTurbo} public numerical program\,\cite{WEBLINK}, following the implementation of Ref.~\cite{Neumann:2022lft}, and have been benchmarked against {\ttfamily MCFM}~\cite{Neumann:2022lft,Boughezal:2015ded}, as well as, upon integration
over $q_T$, against the inclusive cross section computed at the same perturbative order~\cite{Baglio:2022wzu}.

The function ${\cal G}$ is singular when $\alpha_S\tL=\pi/\beta_0$ (where $\beta_0$ is the one-loop coefficient of the QCD $\beta$ function)
which corresponds to the region of transverse-momenta
of the order of the Landau pole of the QCD coupling
or $b^{-1}\sim \Lambda_{\mathrm{QCD}}$. This signals
that a truly non-perturbative  region is approached and
perturbative results (including resummed ones) are not reliable. In this region, a model for NP QCD effects is required, which must include a regularization of the singular behavior of the function $\mathcal{G}$. As noted in Refs.~~\cite{Qiu:2000hf,Catani:1996yz,Laenen:2000de,Lustermans:2019plv}, the specific form of this regularization remains somewhat arbitrary. In this work, we adopt the so-called $b_\star$ prescription, originally proposed in Ref.~\cite{Collins:1984kg},
by freezing the $b$ dependence of $\exp{G(\alpha_S , \tilde{L})}$ before it reaches the Landau singularity.

To achieve this, a parameter $b_{\text{max}}$ is introduced with the replacement
\begin{equation}
\label{e:bstar}
b \to b_\star = \frac{b}{\sqrt{1 + \frac{b^2}{b_{\text{max}}^2}}}\,.
\end{equation}
This prescription guarantees that the variable $b_\star$ saturates to $b_{\text{max}}$ at large values of $b$.
However, $b_{\star}$ also introduces spurious power corrections
that scale like
$(\Lambda_{\text{QCD}}/q_T)^k$~\cite{Catani:1996yz,Kulesza:2002rh,Laenen:2000de,Kulesza:2003wn},
with $k>0$. In the region $q_T \simeq \Lambda_{\text{QCD}}$, these power
corrections become sizeable and have to be modelled by including in
Eq.~\eqref{resum} a non-perturbative form factor.

We incorporate NP effects through a NP form factor of the form  $\exp\{{\cal G}_{NP}(b,Q/Q_0)\}$\,\cite{Catani:2015vma}
which multiplies the perturbative form factor $\exp\{{\cal G}(\alpha_S,\tL)\}$ appearing in Eq.~\eqref{resum}.
We adopt a NP form factor inspired by the model proposed in Ref.~\cite{Collins:2014jpa}, with the following functional form:
\begin{equation}
  \label{gNP}
{\cal G}_{NP}(b,M^2/Q_0^2) = \exp \Big{[} -g_j(b) - g_K(b) \ \log \frac{M^2}{Q_0^2} \Big{]}\, .
\end{equation}

The function $g_j(b)$
accounts for the non-perturbative part of the TMDs, or, more generally,
for the intrinsic transverse momentum of the parton inside the
parent hadron, at $M=Q_0$ We have used the follwing parameterization
for $g_j(b)$ from Ref.\,\cite{ATLAS:2023lhg}
\begin{equation}
\label{gjNP}
    g_j(b) = \frac{g_1 b^2}{1+\lambda b^2} + \text{sign}(q) \bigg{(}1 - \exp[-|q| b^4] \bigg{)}\,.
\end{equation}
We note that we have not included an additional flavour dependence in $g_j(b)$, because we have obtained a good description of the experimental data even without it.
We also observe that recent studies~\cite{Bacchetta:2024qre,Billis:2024dqq,Moos:2023yfa} have found that Drell--Yan data alone are not sufficient to constrain flavour dependence of such non-perturbative effects.

The function $g_K(b)$, instead,
takes into account for the non-perturbative contribution to the Collins--Soper kernel\,\cite{Collins:2011zzd,Collins:1981uk},
that is, the $M^2$ dependence of the TMDs
(see Sec.~\ref{sec:CSK} for further details).

We have chosen to adopt the following parameterization of $g_K(b)$
 proposed in Ref.~\cite{Collins:2014jpa}
 \begin{equation}
\label{gKNP}
    g_K(b) = g_0 \Big{(} 1 - \exp \Big{[} - \frac{C_F \, \alpha_S \, (\mu_{b_\star}^2) b^2}{\pi g_0 b^2_{max}} \Big{]} \Big{)}\,.
\end{equation}
Such parameterization fulfills the following criteria: it does not affect the perturbative calculation at small $b_T$, it saturates to a finite value at large $b_T$, and in the intermediate region it either does not depend or only weakly depends on the choice of $b_{\text{max}}$. This property is particularly important, since $b_{\text{max}}$ itself represents a (hyper)parameter that must be carefully controlled in the fit.

\section{Numerical results}

In the following, we present a comparison between experimental data and theoretical predictions for $Z/\gamma^*$ production and leptonic decay.
The predictions are obtained at central scales, i.e., $\mu_R = \mu_F = Q = M$, and at N$^4$LL+N$^4$LO\,\footnote{We have checked that the use of N$^3$LL+N$^3$LO accuracy has a very small impact on the results presented.} approximated accuracy \cite{Camarda:2023dqn} by matching the resummed calculation with fixed-order results at large $q_T$, up to
NNLO.
The hadronic cross section is computed by convoluting the partonic cross section in Eq.\eqref{resum} with the PDFs from the MSHT set\cite{McGowan:2022nag} at N$^3$LO, using $\alpha_s(m_Z^2) = 0.118$.
For the electroweak couplings, we adopt the so-called $G_\mu$ scheme, with the corresponding input parameters listed in Table~\ref{t_EW_inputs}.

\begin{table}[h!]
\centering
\begin{tabular}{|c|c|c|}
\hline
\textbf{Constant} & \textbf{Value} & \textbf{Unit} \\
\hline
$G_F$       & $1.1663787 \times 10^{-5}$ & GeV$^{-2}$ \\
$m_Z$       & $91.1876$                  & GeV        \\
$\Gamma_Z$  & $2.4950$                   & GeV        \\
$m_W$       & $80.385$                   & GeV        \\
\hline
\end{tabular}
\caption{Numerical values of the electroweak constants used as input for the $G_\mu$ scheme in our calculations.}
\label{t_EW_inputs}
\end{table}

Our calculation includes the leptonic decays $Z/\gamma^* \to \ell^+ \ell^-$, taking into account the effects of $Z/\gamma^*$ interference and the finite width $\Gamma_Z$ of the $Z$ boson. Even if the measurements considered in the analysis are inclusive in the lepton kinematics, our calculation can retain spin correlations and the full dependence on the kinematical variables of the final-state leptons.

The datasets considered in our numerical analysis are listed in Tab.~\ref{t:data}. Although this is not a global analysis including all available data, it encompasses both high- and low-energy measurements, originating from collider experiments (ATLAS and CDF) and fixed-target experiments (E605 and E288), respectively. The selected data span a broad kinematic range and provide sensitivity to different regions of phase space. Moreover, as shown in the table, the datasets differ in their coverage and structure: some are subject to cuts in rapidity $y$, others in Feynman-$x$ ($x_F$) variable, and some are fully inclusive. The  selected dataset has been chosen to
be sufficiently representative to demonstrate the robustness of our  results. In total, we analyze 378 data points.
For clarity, we note that in this work we include the majority of the available data points in the $q_T$ spectra for the low invariant mass region, i.e. from fixed-target experiments, and for the high-energy datasets we include data up to $q_T < 30$~GeV. This choice is motivated by our primary interest in describing the low invariant mass region, together with the fact that the applicability of the this formalism to high-energy data has already been extensively validated in the literature.

\begin{table}[h!]
\footnotesize
\begin{center}
\renewcommand{\tabcolsep}{0.4pc}
\renewcommand{\arraystretch}{1.2}
\begin{tabular}{|c|c|c|c|c|c|c|}
  \hline
  Experiment & $N_{\rm dat}$ & Observable  &  $\sqrt{s}$ [GeV]& $Q$ [GeV] &  $y$ or $x_F$ & Ref. \\
  \hline
  \hline
  E605 & 74 & $E d^3\sigma/d^3 \bm{q}$ & 38.8 & 7 - 18  & $x_F=0.1$  & \cite{Moreno:1990sf} \\
  \hline
  E288 200 GeV & 50 & $E d^3\sigma/d^3 \bm{q}$ &  19.4  & 4 - 9  & $y=0.40$ & \cite{Ito:1980ev} \\
  \hline
  E288 300 GeV & 59 & $E d^3\sigma/d^3 \bm{q}$ &  23.8  & 4 - 12  & $y=0.21$ & \cite{Ito:1980ev} \\
  \hline
  E288 400 GeV & 68 & $E d^3\sigma/d^3 \bm{q}$ &  27.4  & 5 - 14  & $y=0.03$ & \cite{Ito:1980ev} \\
  \hline
  CDF Run II & 55 & $d\sigma/d q_T$ & 1960 & 66 - 116  &  Inclusive & \cite{CDF:2012brb} \\
  \hline
  \makecell{ATLAS 8 TeV} & \makecell{9\\9\\9\\9\\9\\9\\9\\9} & $(1/\sigma)d\sigma/d q_T$ & 8000 & 80 - 100  & \makecell{$0<|y|<0.4$ \\ $0.4<|y|<0.8$ \\ $0.8<|y|<1.2$\\$1.2<|y|<1.6$\\$1.6<|y|<2.0$\\$2.0<|y|<2.4$\\$2.4<|y|<2.8$\\$2.8<|y|<3.6$} & \cite{ATLAS:2023lsr} \\
  \hline
  \hline
  Total & 378 & & & & & \\
  \hline
\end{tabular}
\caption{Experimental data sets included in this analysis. Each row contains the number of data points ($N_{\rm dat}$), the measured observable, the center-of-mass energy$\sqrt{s}$, the invariant mass range, the acceptance region in the
variable $y$ or $x_F$, and the published reference.}
\label{t:data}
\end{center}
\end{table}

To determine the optimal values of the nonperturbative parameters entering Eqs.~\eqref{gNP}-\eqref{gKNP}, we performed a fit by interfacing {\tt DYTurbo} with {\tt xFitter}~\cite{Alekhin:2014irh}. In particular, the best-fit parameters were obtained by minimizing a $\chi^2$ function that accounts for both experimental and theoretical uncertainties, the latter arising from PDF variations. The $\chi^2$ function is defined as
\begin{equation}
\chi^2(\beta_\text{exp}, \beta_\text{th}) = \sum_{i=1}^{N_{\text{dat}}} \frac{\left( \sigma_i^{\text{exp}} + \sum_j \Gamma_{ij}^{\text{exp}} \beta_{j,\text{exp}} - \sigma_i^{\text{th}} - \sum_k \Gamma_{ik}^{\text{th}} \beta_{k,\text{th}} \right)^2}{\Delta_i^2} + \sum_j \beta_{j,\text{exp}}^2 + \sum_k \beta_{k,\text{th}}^2 \,,
\label{e:chi2}
\end{equation}
where the index $i$ runs over all $N_{\text{dat}}$ data points, while $j$ and $k$ run over the nuisance parameter vectors $\beta_{\text{exp}}$ and $\beta_{\text{th}}$, which encode the correlated experimental and theoretical (PDF-related) uncertainties, respectively. The effect of each of them on the experimental data and theoretical prediction is described by the response matrices $\Gamma_{ij}^{\text{exp}}$ and $\Gamma_{ik}^{\text{th}}$. The measurements and the uncorrelated experimental uncertainties are denoted by $\sigma_i^{\text{exp}}$ and $\Delta_i$, respectively, while the theoretical predictions are denoted by $\sigma_i^{\text{th}}$.

\begin{figure}[h]
    \centering
\includegraphics[width=0.4\linewidth]{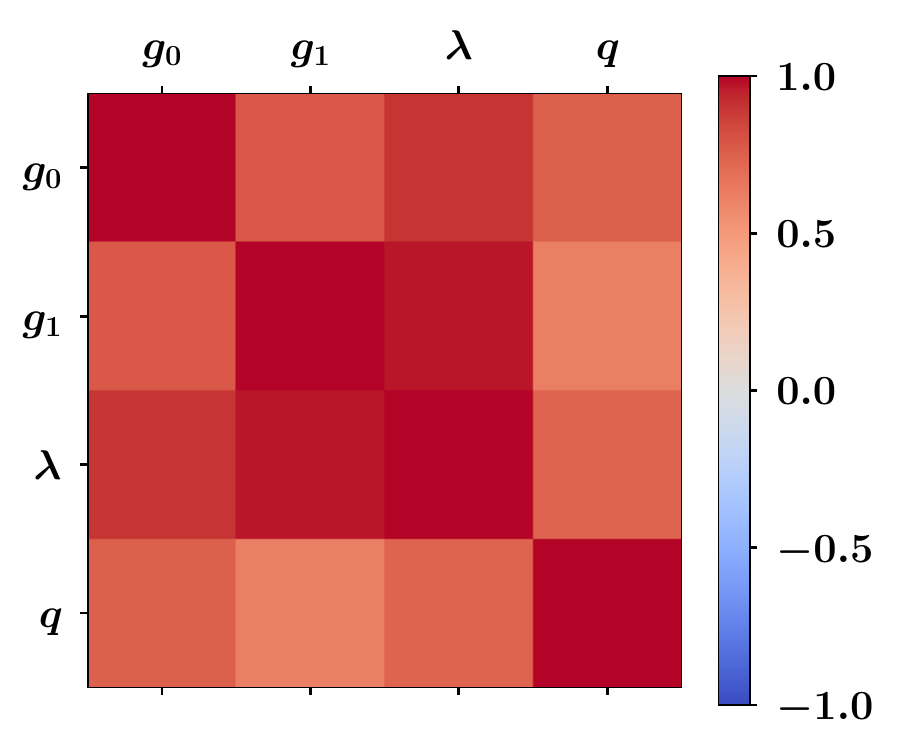}
    \caption{Graphical representation of the correlation matrix for the free parameters of fit.}
    \label{f:Cov_Mat}
\end{figure}

As part of the fitting procedure, we also investigated the dependence of the results on the parameters $Q_0$ and $b_{\text{max}}$ appearing in Eqs.~\eqref{e:bstar}–\eqref{gKNP}, which are typically chosen a priori. We found that the quality of the fit, as reflected in the predictions and quantified by the resulting $\chi^2$, is largely insensitive to variations in these parameters\footnote{This behavior was already suggested in Ref.\cite{Collins:1984kg}, and we have explicitly confirmed it in our analysis.}. Nevertheless, in order to fix their values, we identified the most stable and accurate configuration as $Q_0 = 1$~GeV and $b_{\text{max}} = 1.5$~GeV$^{-1}$. The final fit yields a reduced chi-squared of $\chi^2 / N_{\text{d.o.f.}} = 1.25$, and the corresponding best-fit parameters, along with their uncertainties extracted via the Hessian method, are listed in Tab.~\ref{t:NPpar}. The associated covariance matrix is shown in Fig.~\ref{f:Cov_Mat}.
\begin{table}[h]
    \centering
    \begin{tabular}{|c|c|c|c|c|}
    \hline
           & $g_0$ & $g_1$ & $\lambda$ & $q$ \\
         \hline
         NP Parameters & $0.83 \pm 0.14$ & $0.732 \pm 0.046$ & $0.92 \pm 0.20$ & $-0.136 \pm 0.010$ \\
         \hline
    \end{tabular}
    \caption{Values obtained for the free parameters in the fit. For each parameter, the best-fit value and the associated standard deviation are reported.}
    \label{t:NPpar}
\end{table}

As we can see from the table, all parameters are reasonably well constrained, although some exhibit significant mutual correlations. Together with the value $\chi^2 / N_{\text{d.o.f.}} = 1.25$~\footnote{We performed an additional fit using the NNPDF40an3lo Hessian set~\cite{NNPDF:2024nan}, keeping the same configuration as before, and obtained a $\chi^2 / N_{\text{d.o.f.}} = 1.38$. The deterioration in the fit quality is likely due to the smaller PDF uncertainties associated with this set.}, this suggests that the chosen functional form for the non-perturbative component, which involves only a small number of parameters, is capable of providing a good description of the data. The presence of strong correlations is a common feature in fits of hadronic distributions and may indicate that an alternative functional form could improve the description. However, a systematic exploration of this possibility lies beyond the scope of the present work and is left for future studies.

Before presenting the comparison between the data and the predictions obtained from the fit, we emphasize that we performed several fits by varying the set of included data points, particularly by applying different cuts on the ratio $\frac{q_T}{Q}$. We observed that including a larger number of data points generally leads to tighter constraints on the non-perturbative parameters, especially $\lambda$ and $q$. In particular, for the fit performed with the cut $\frac{q_T}{Q} \leq 0.2$, which includes 240 data points and corresponds to the typical kinematic range considered in standard TMD fits~\cite{Bacchetta:2024qre,Bacchetta:2025ara,Moos:2025sal,Moos:2023yfa,Moos:2025sal,Bacchetta:2022awv}, we obtain a reduced chi-squared of $\chi^2 / N_{\text{d.o.f.}} = 1.03$. This result is comparable to, or even better than, those from existing TMD fits, despite relying on a significantly smaller number of non-perturbative parameters. The quality of the description slightly worsens when data at higher $q_T$ are included, which may be attributed either to statistical issues in the low-energy data or to missing-mass effects and higher-order corrections in the fixed-order contributions, which become more relevant at low invariant masses.

\begin{figure}[h!]
\centering
\subfigure{\includegraphics[width=0.49\textwidth]{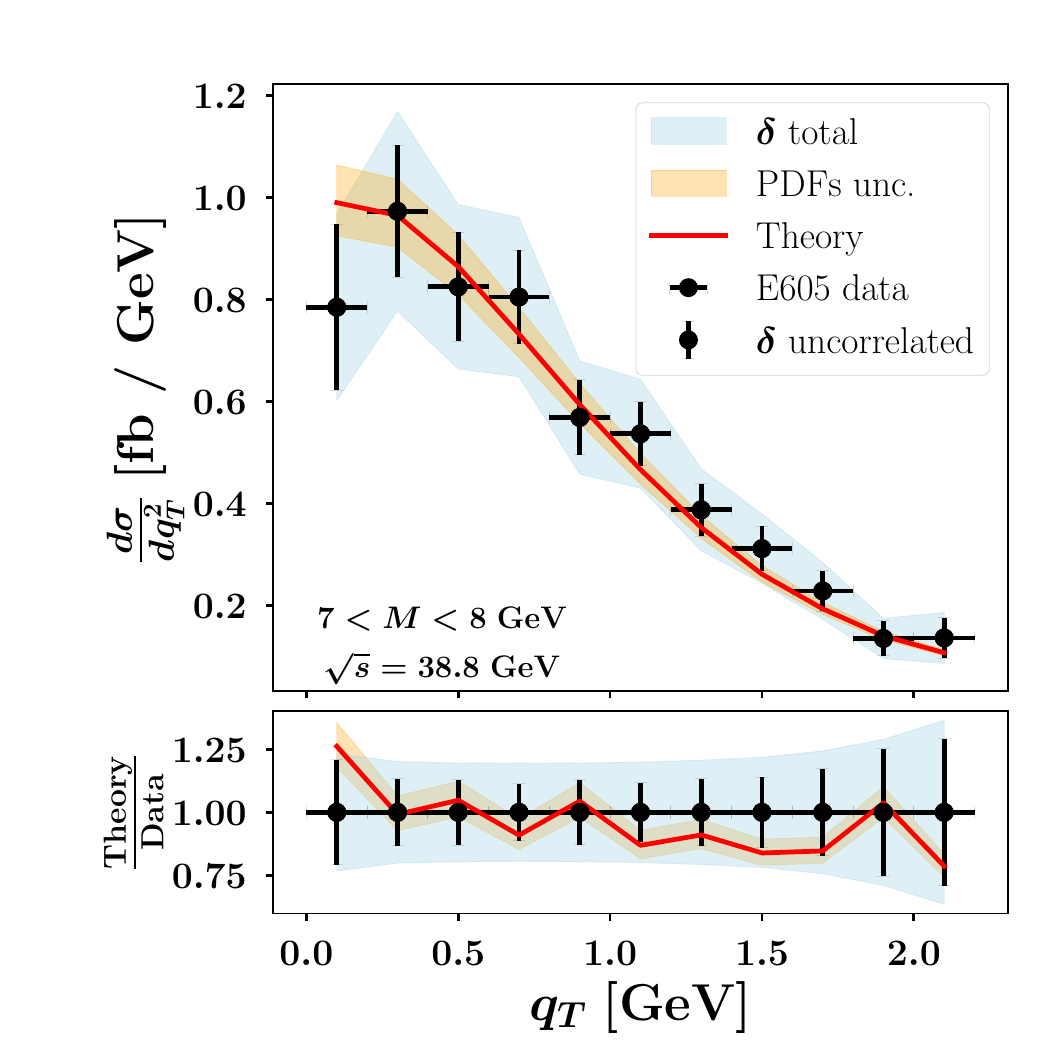}}
\hfill
\subfigure{\includegraphics[width=0.49\textwidth]{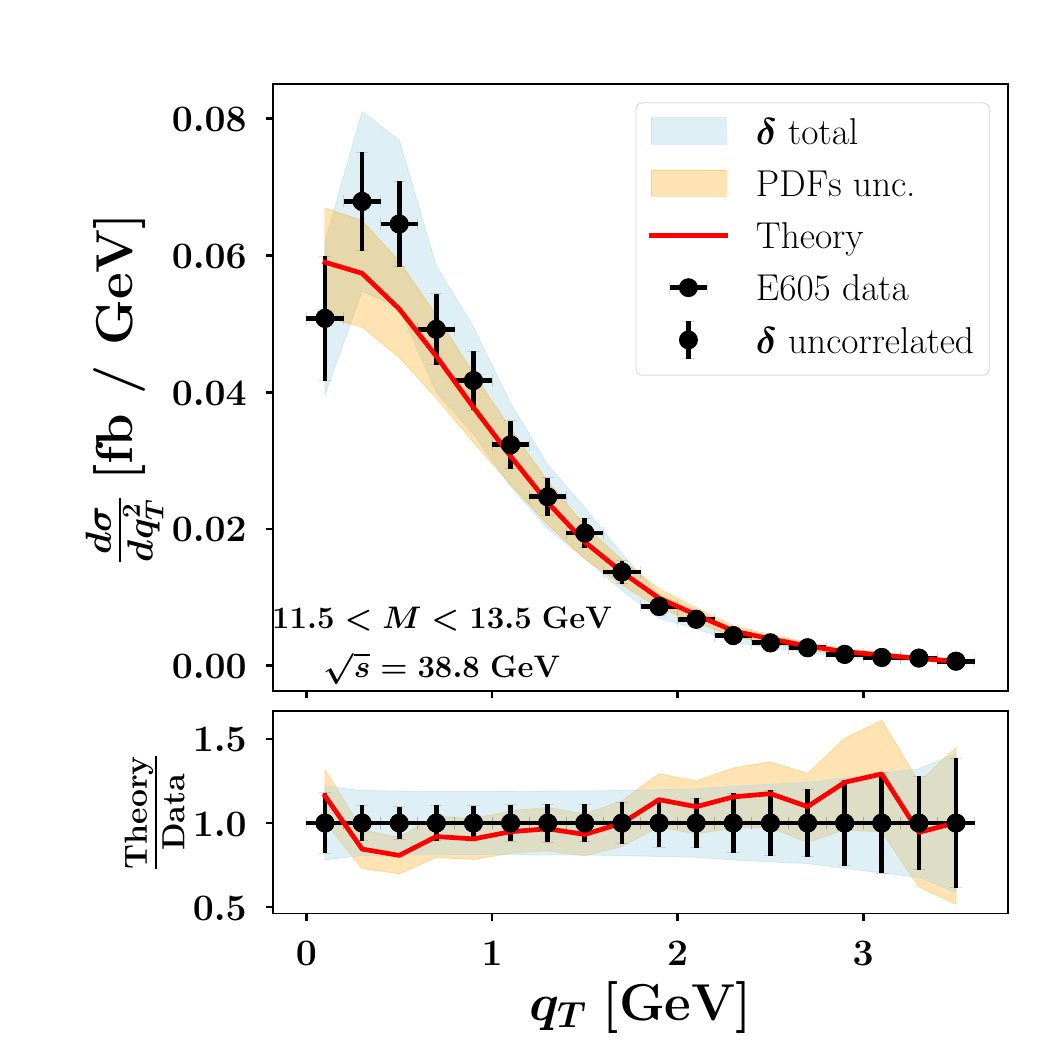}}
\caption{Comparison between theoretical predictions and experimental data for the Drell–Yan cross section differential in $q_T$ from the E605 experiment. The left and right panels correspond to the invariant mass bins $7 < M < 8$~GeV and $11.5 < M < 13.5$~GeV, respectively. The orange bands represent the PDF uncertainty from the MSHT20an3lo set.}
\label{f:E605}
\end{figure}

In Fig.~\ref{f:E605}, we compare the predictions obtained from the fit with the experimental data from the E605 experiment for the invariant mass bins $7 < M < 8$~GeV and $11.5 < M < 13.5$ GeV. The orange bands represent the PDF uncertainty from the MSHT20an3lo set. As shown, the experimental data are well described within uncertainties across the entire transverse momentum spectrum. This provides clear evidence that the formalism described by Eq.~\eqref{partXS2}, supplemented by a simple non-perturbative model, is capable of accurately describing low invariant mass Drell–Yan data.

\begin{figure}[h!]
\centering
\subfigure{\includegraphics[width=0.325\textwidth]{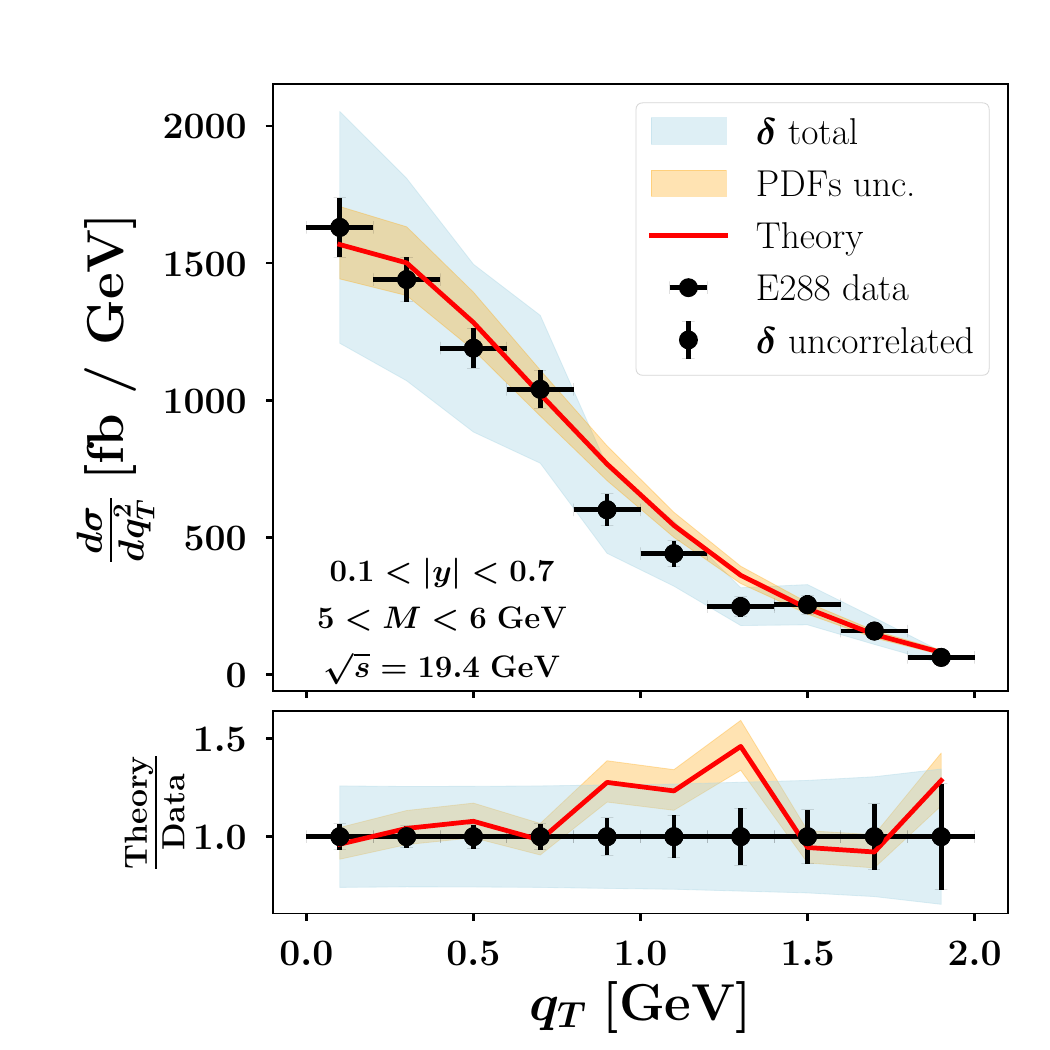}}
\hfill
\subfigure{\includegraphics[width=0.325\textwidth]{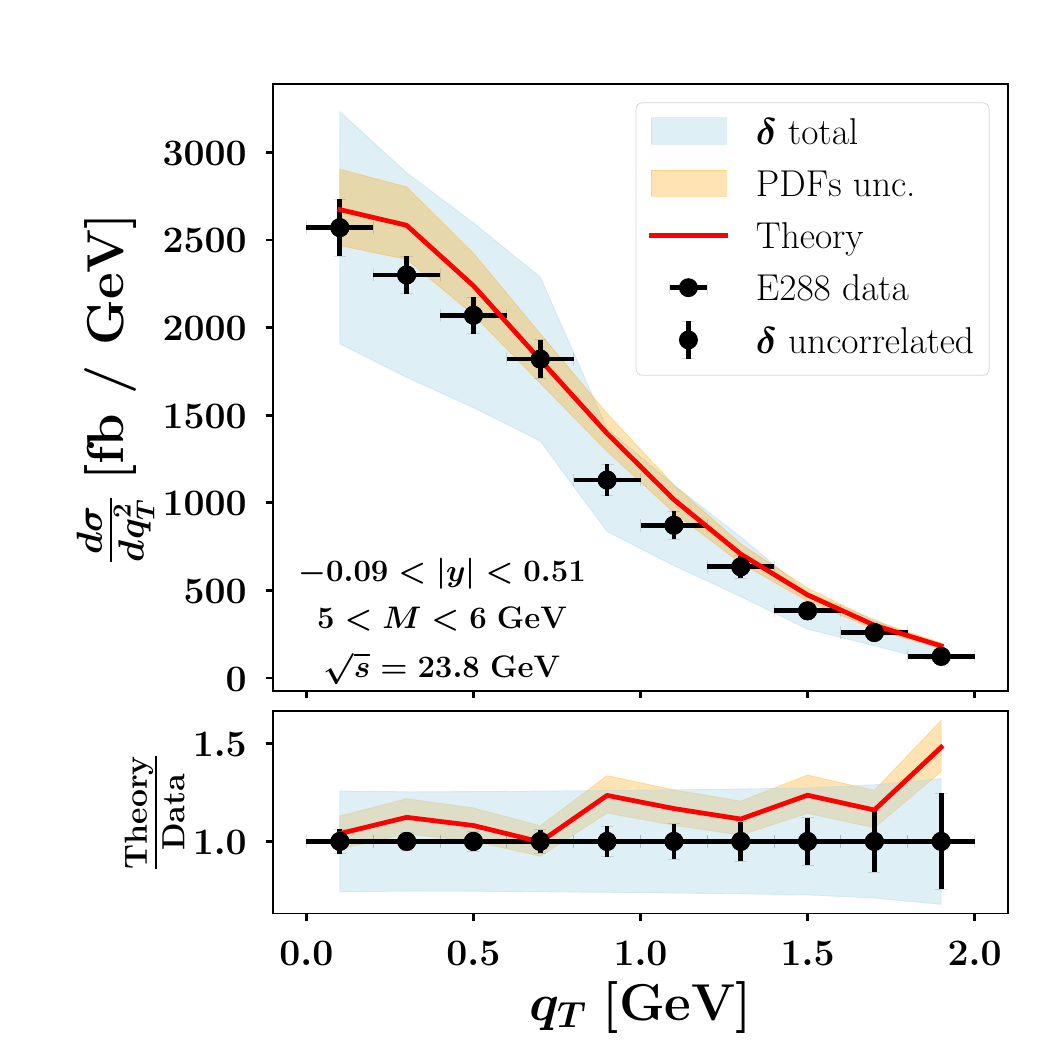}}
\hfill
\subfigure{\includegraphics[width=0.325\textwidth]{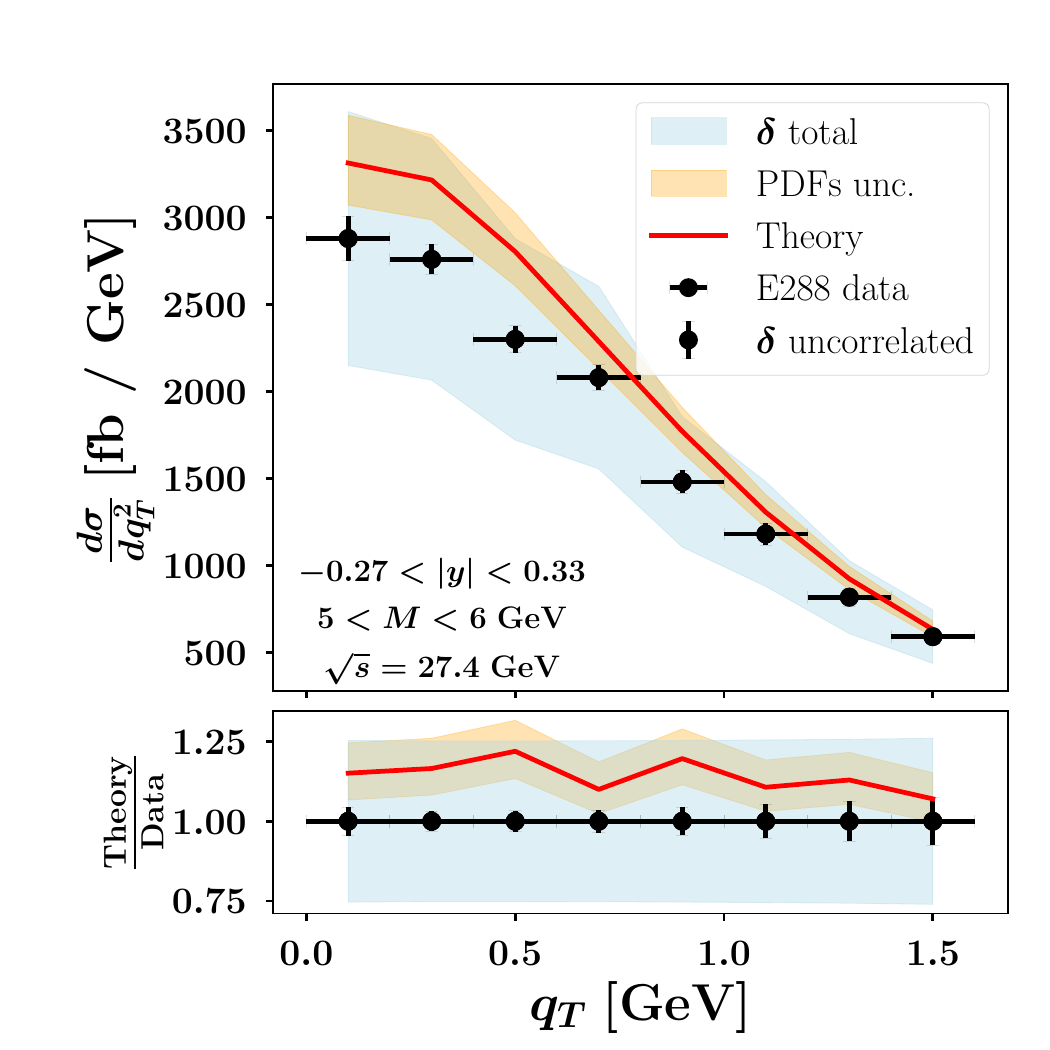}}
\caption{Comparison between theoretical predictions and experimental data for the Drell–Yan cross section differential in $q_T$ from the E288 experiment, in the invariant mass bin $5 < M < 6$~GeV. The left, center, and right panels correspond to beam energies of $E_{\text{beam}} = 200$~GeV, $300$~GeV, and $400$~GeV, respectively. The orange bands represent the PDF uncertainty from the MSHT20an3lo set.}
\label{f:E288_5_6}
\end{figure}

\begin{figure}[h]
\centering
\subfigure{\includegraphics[width=0.325\textwidth]{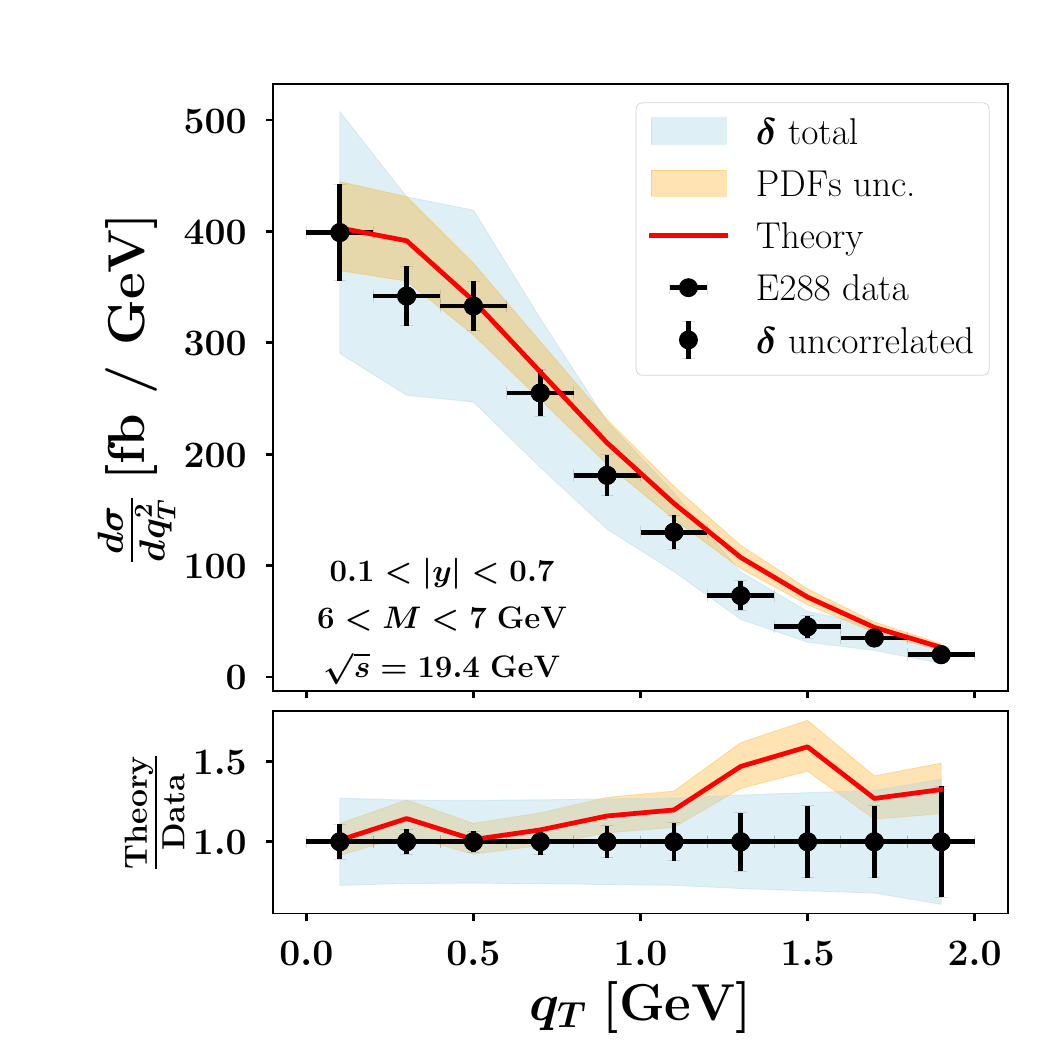}}
\hfill
\subfigure{\includegraphics[width=0.325\textwidth]{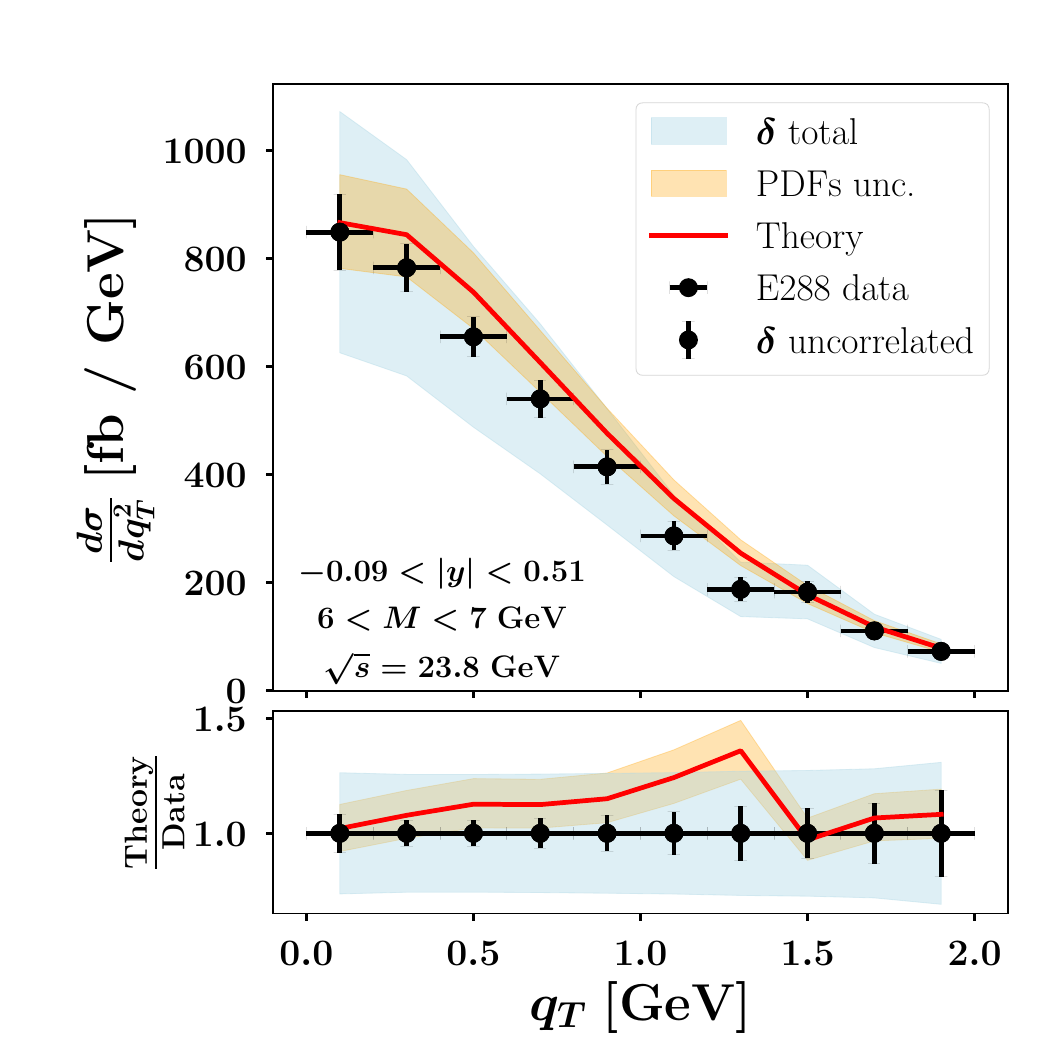}}
\hfill
\subfigure{\includegraphics[width=0.325\textwidth]{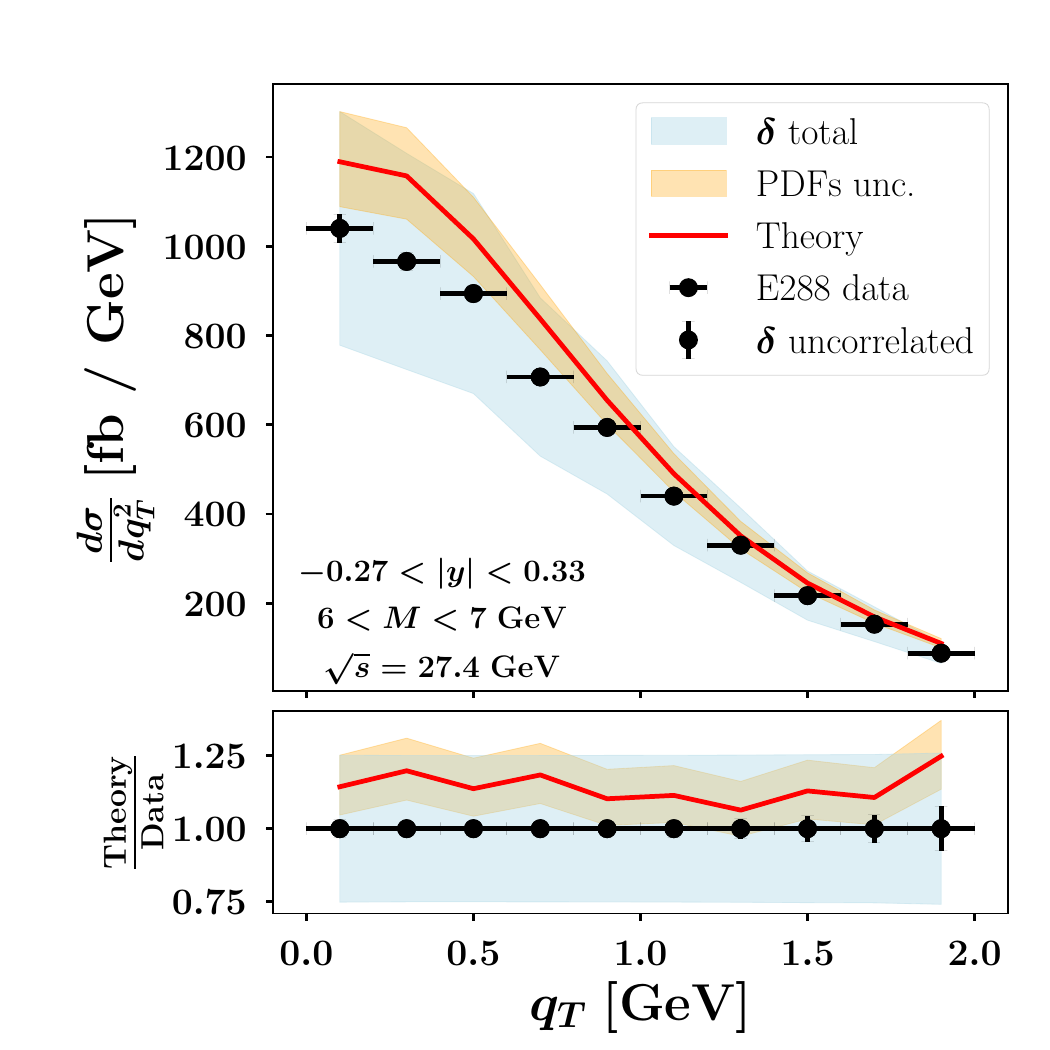}}
\caption{Comparison between theoretical predictions and experimental data for the Drell–Yan cross section differential in $q_T$ from the E288 experiment, in the invariant mass bin $6 < M < 7$~GeV. The left, center, and right panels correspond to beam energies of $E_{\text{beam}} = 200$~GeV, $300$~GeV, and $400$~GeV, respectively. The orange bands represent the PDF uncertainty from the MSHT20an3lo set.}
\label{f:E288_6_7}
\end{figure}

Continuing the analysis, Figs.~\ref{f:E288_5_6} and~\ref{f:E288_6_7} show the comparison between theoretical predictions and data from the E288 experiment for the invariant mass bins $5 < M < 6$~GeV and $6 < M < 7$~GeV, respectively, at three different beam energies: $E_{\text{beam}} = 200$, $300$, and $400$~GeV. In all cases, we observe very good agreement within the combined experimental and theoretical uncertainties. A slight deterioration in the description is observed in the highest $q_T$ bins, which may indicate the presence of missing higher-order corrections or target mass effects, both of which are neglected in the present analysis.

\begin{figure}[h!]
\centering
\subfigure{\includegraphics[width=0.49\textwidth]{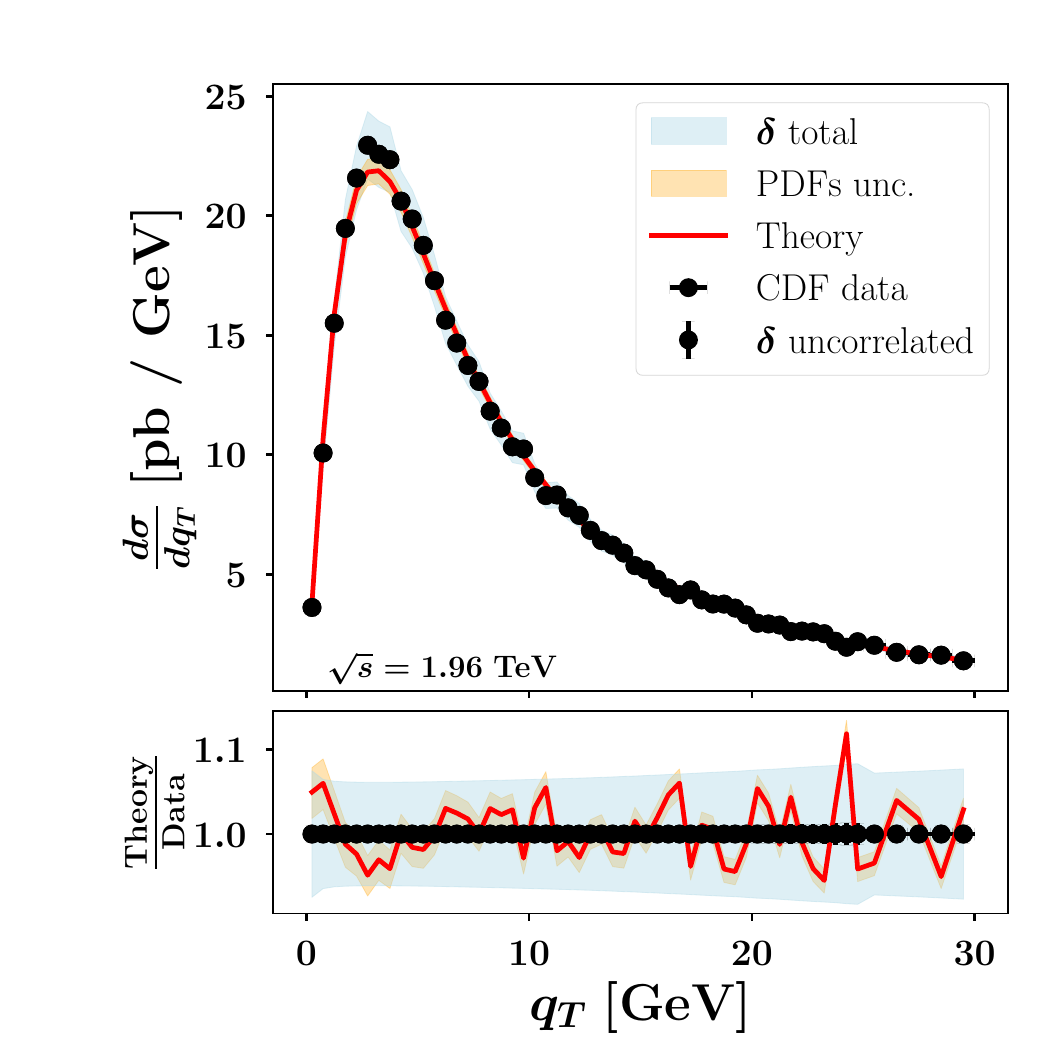}}
\hfill
\subfigure{\includegraphics[width=0.49\textwidth]{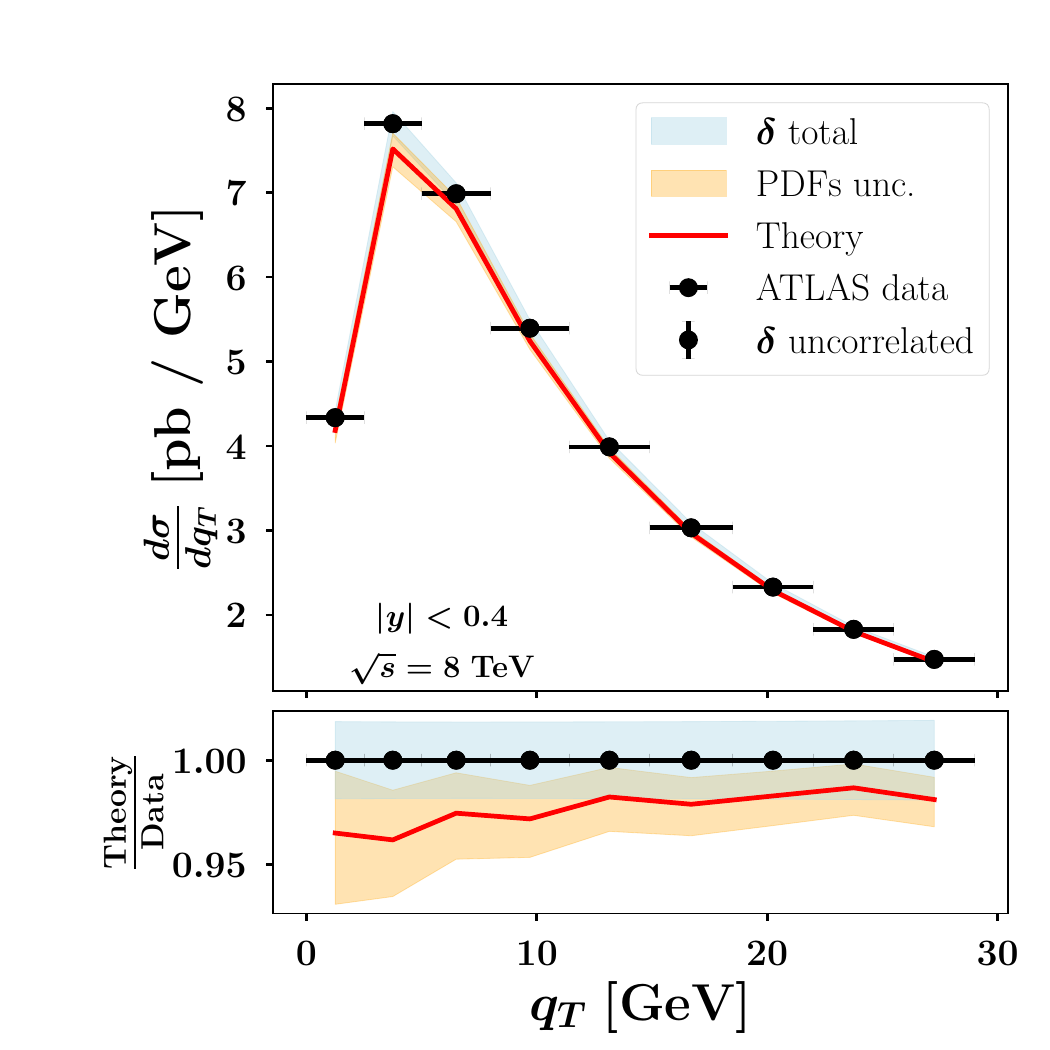}}
\caption{Comparison between theoretical predictions and high-energy Drell–Yan data at the $Z$-boson peak. The left panel corresponds to the CDF measurement, while the right panel shows the ATLAS data at $\sqrt{s} = 8$~TeV in the central rapidity region $|y| < 0.4$.}
\label{f:Zboson}
\end{figure}

For completeness, in Fig.~\ref{f:Zboson} we also show the comparison between theoretical predictions and high-energy Drell–Yan data at the $Z$-boson peak. It is well established in the literature that these data are accurately described by perturbative QCD combined with a simple non-perturbative model; see, for instance, Refs.\cite{Catani:2015vma}. The left panel shows the result for the CDF experiment, while the right panel corresponds to the ATLAS measurement at $\sqrt{s} = 8$~TeV in the central rapidity region $|y| < 0.4$. We observe excellent agreement with the experimental data within the uncertainty bands. Importantly, we find that the same theoretical framework and non-perturbative model used for the low-energy data also provides a consistent and accurate description of the high-energy regime. This confirms the robustness and flexibility of our approach in describing the full range of available Drell–Yan data across different energies.

\subsection{Determination of the Collins-Soper Kernel}
\label{sec:CSK}
In recent years, the Collins–Soper kernel has attracted considerable attention, as it plays a crucial role in the TMD formalism by governing the evolution of TMDs with respect to the rapidity scale $\zeta$\,\cite{Collins:1981uk,Collins:2011zzd,Echevarria:2011epo}. Its relevance is further underscored by the fact that it can be directly extracted from experimental data~\cite{Bacchetta:2024qre,Bacchetta:2025ara,Moos:2025sal,Moos:2023yfa,Bacchetta:2022awv} or estimated using lattice QCD calculations~\cite{LatticePartonLPC:2023pdv,Avkhadiev:2023poz,Avkhadiev:2024mgd,Bollweg:2024zet,Bollweg:2025iol,LatticePartonLPC:2022eev}.\\

The Collins–Soper kernel is related to the cusp
anomalous dimension by the following equation~\cite{Collins:1984kg}
\begin{equation}
    \frac{dK(b^2Q^2,\alpha_S(Q^2)))}{d\ln Q^2} = - \gamma_K(\alpha_S(Q^2))\, ,
\end{equation}
where $\gamma_K(\alpha_S)$ is the cusp anomalous dimension\,. The solution of the above reads
\begin{equation}
    K(b^2Q^2,\alpha_S(Q^2)) = K(b_0^2,\alpha_S(b_0^2/b^2)) - \int_{b_0^2/b^2}^{Q^2} \frac{dq^2}{q^2} \gamma_K(\alpha_S(q^2))\,.
\end{equation}
The kernel can be perturbatively expanded as
\begin{align}
K(b^2Q^2,\alpha_S) &= \sum_{n=1}^{\infty} \left( \frac{\alpha_S}{\pi} \right)^{n} K^{(n)}(b^2Q^2) \,.
\end{align}

The Collins--Soper kernel can be related
to the functions $A(\alpha_S)$ and $B(\alpha_S)$ in Eq.~\eqref{e:Gfactor},
through the relations~\cite{Collins:1984kg}:
\begin{align}
A(\alpha_S) &= \frac{1}{2} \gamma_K(\alpha_S) + \frac{1}{2} \beta(\alpha_S) \frac{\partial  K(b_0^2, \alpha_S)}{\partial  \ln\alpha_S} \, , \\
B(\alpha_S) &= -\gamma_F(\alpha_S)
- K(b_0^2,\alpha_S) \, ,
\end{align}
and by inverting these relations one can obtain the correspondence between the perturbative coefficients $K^{(n)}, \gamma_K^{(n)}$ and $A^{(n)}$, $B^{(n)}$, $\gamma_F^{(n)}$, which arise respectively from the perturbative expansions of the $A(\alpha_S)$ and $B(\alpha_S)$ functions and from the non-cusp anomalous dimension $\gamma_F(\alpha_S)$\footnote{In Ref.~\cite{Collins:1984kg}, $\gamma_F(\alpha_S)$ was called $G(\alpha_S)$.}.
\\
The  non-perturbative contribution of the Collins--Soper kernel is encoded in the function $g_K(b)$ in Eq.\,(\ref{gKNP}).

\begin{figure}[h]
\centering
\includegraphics[width=0.9\linewidth]{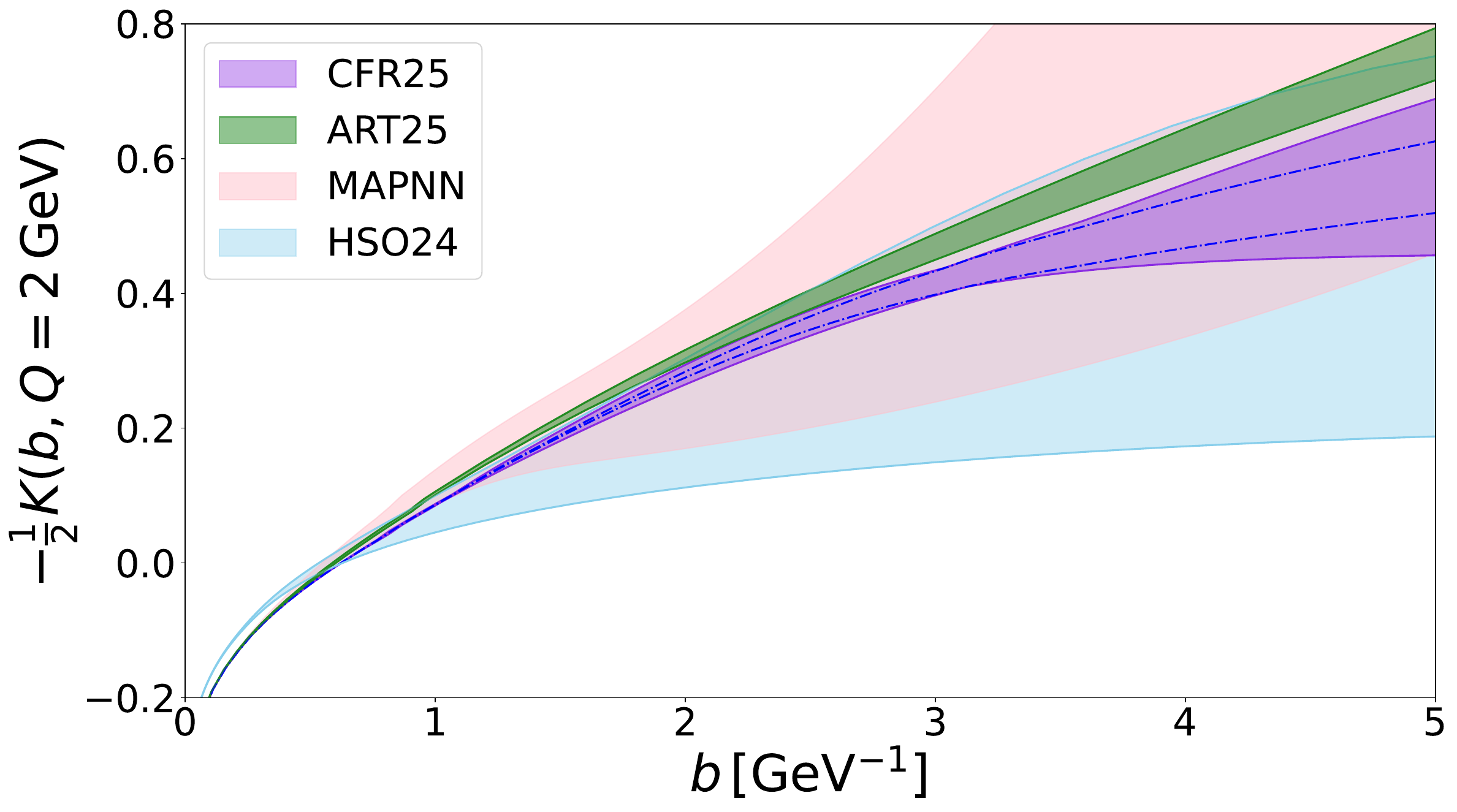}
\caption{The Collins–Soper kernel as a function of $b$ at the scale $\mu_R = Q = 2$~GeV, as extracted in the present analysis (CFR25), compared to other determinations MAPNN~\cite{Bacchetta:2025ara}, ART25~\cite{Moos:2025sal}, HSO24~\cite{Aslan:2024nqg}. The purple band represents the combined uncertainties from the parameters $g_0$, $b_{\text{max}}$, and $Q_0$, while the blue dot-dashed lines indicate the uncertainty due to $g_0$ only.}
\label{f:CSK}
\end{figure}

In Fig.~\ref{f:CSK} we show the perturbative and non-perturbative contribution to the Collins--Soper kernel as a function of $b$:
\begin{equation}
K(b,Q) = K(b_\star^2Q^2,\alpha_S) + g_K(b)\,.
\end{equation}
with the scale fixed, as conventionally  in the literature, at $Q = 2$~GeV. The purple uncertainty band accounts for the combined effect of the uncertainty on the fitted parameter $g_0$, as well as the variations of $b_{\text{max}}$ in the range $1$–$2.5$~GeV$^{-1}$ and $Q_0$ in the range $1$–$3$~GeV. For comparison, the blue dot-dashed lines indicate the uncertainty due to $g_0$ only. Notably, it also accounts for the uncertainties associated with the choice of $b_{\text{max}}$, which is often fixed a priori in other analyses, which potentially introduces a theoretical bias.

 In Fig.~\ref{f:CSK}, we also show a comparison
of our extraction of the Collins–Soper kernel against the ones
of other groups. Our result is found to be compatible with the latest determination by the MAP Collaboration, obtained from fits to low-transverse-momentum Drell--Yan data~\cite{Bacchetta:2025ara}, as well as with the extraction of Ref.~\cite{Aslan:2024nqg}, where low-invariant-mass Drell--Yan data were analyzed. It is also in reasonable agreement with the most recent extraction by the ART Collaboration, which is based on combined low-transverse-momentum Drell--Yan and SIDIS data~\cite{Moos:2025sal}. A slight tension between the uncertainty bands can be observed in this case; however, we note that the authors of Ref.~\cite{Moos:2025sal} explicitly acknowledge in Ref.~\cite{Cuerpo:2025zde} that their uncertainty estimates may be somewhat underestimated.

As discussed in the previous section, a more accurate determination of the nonperturbative parameters is obtained when including higher values of $q_T$ than those typically used in TMD fits. Therefore, the most reliable extractions of the Collins–Soper kernel should rely on global analyses that incorporate all available experimental data and exploit the full region in which resummation effects have a non-negligible impact, as emphasized in the present study. A more detailed and dedicated investigation of the Collins–Soper kernel along these lines is left for future work.

\section{Conclusions}

In this work, we have presented a comprehensive phenomenological analysis of the Drell-Yan transverse-momentum ($q_T$) distribution, covering a broad kinematic range from low invariant masses up to the $Z$-boson peak ($4\leq M \leq 116$~GeV). Our predictions include the state-of-the-art ingredients  in perturbative QCD: we combine the resummation of large Sudakov logarithms at N$^4$LL accuracy with N$^4$LO hard-virtual contributions at small $q_T$ and we consistently match resummed results with the known  $\mathcal{O}(\alpha_S^3)$ fixed-order calculation valid at large $q_T$.

These contributions have been implemented in the {\ttfamily DYTurbo} public numerical program\,\cite{WEBLINK}
which allows the user to apply arbitrary kinematical cuts on the vector boson and the final-state leptons, and
which provides fast and
numerically precise predictions
for the  relevant kinematical distributions.

We have found that, within our resummation formalism of Refs.\,\cite{Catani:2000vq,Bozzi:2005wk,Bozzi:2007pn},  pure perturbative QCD predictions are able to describe, within the
errors, experimental data down to $q_T \sim 1$~GeV.
At very low values of $q_T$ ($q_T \sim 1$~GeV) non-perturbative (NP) QCD effects are essential and have been included  through a NP  form factor
depending on very few free parameters.

We performed a detailed comparison of our predictions with a wide array of experimental data from fixed-target  and collider experiments. We find excellent agreement between our theoretical framework and the data across all energies and $q_T$ ranges analyzed.

The remarkable consistency between theory and data has enabled a precise extraction of the parameters governing the NP form factor and the so-called Collins-Soper kernel.

\paragraph{Acknowledgments.}
The work of L.R. is partially supported by the Italian Ministero dell'Università e Ricerca (MUR) through the research grant 20229KEFAM (PRIN2022, Next Generation EU, CUP H53D23000980006).

\bibliographystyle{JHEP}
\bibliography{DYqT-lowQ2}

@article{Camarda:2021ict,
    author = "Camarda, Stefano and Cieri, Leandro and Ferrera, Giancarlo",
    title = "{Drell{\textendash}Yan lepton-pair production: qT resummation at N3LL accuracy and fiducial cross sections at N3LO}",
    eprint = "2103.04974",
    archivePrefix = "arXiv",
    primaryClass = "hep-ph",
    doi = "10.1103/PhysRevD.104.L111503",
    journal = "Phys. Rev. D",
    volume = "104",
    number = "11",
    pages = "L111503",
    year = "2021"
}

@article{Ebert:2020dfc,
    author = "Ebert, Markus A. and Michel, Johannes K. L. and Stewart, Iain W. and Tackmann, Frank J.",
    title = "{Drell-Yan $q_{T}$ resummation of fiducial power corrections at N$^{3}$LL}",
    eprint = "2006.11382",
    archivePrefix = "arXiv",
    primaryClass = "hep-ph",
    reportNumber = "DESY-20-016, DESY 20-016, MIT-CTP 5205",
    doi = "10.1007/JHEP04(2021)102",
    journal = "JHEP",
    volume = "04",
    pages = "102",
    year = "2021"
}

@article{Catani:2015vma,
    author = "Catani, Stefano and de Florian, Daniel and Ferrera, Giancarlo and Grazzini, Massimiliano",
    title = "{Vector boson production at hadron colliders: transverse-momentum resummation and leptonic decay}",
    eprint = "1507.06937",
    archivePrefix = "arXiv",
    primaryClass = "hep-ph",
    reportNumber = "ICAS-02-15, TIF-UNIMI-2015-10, ZU-TH-24-15",
    doi = "10.1007/JHEP12(2015)047",
    journal = "JHEP",
    volume = "12",
    pages = "047",
    year = "2015"
}

@article{Catani:2012qa,
    author = "Catani, Stefano and Cieri, Leandro and de Florian, Daniel and Ferrera, Giancarlo and Grazzini, Massimiliano",
    title = "{Vector boson production at hadron colliders: hard-collinear coefficients at the NNLO}",
    eprint = "1209.0158",
    archivePrefix = "arXiv",
    primaryClass = "hep-ph",
    reportNumber = "ZU-TH-16-12",
    doi = "10.1140/epjc/s10052-012-2195-7",
    journal = "Eur. Phys. J. C",
    volume = "72",
    pages = "2195",
    year = "2012"
}

@article{Bozzi:2003jy,
    author = "Bozzi, G. and Catani, S. and de Florian, D. and Grazzini, M.",
    title = "{The q(T) spectrum of the Higgs boson at the LHC in QCD perturbation theory}",
    eprint = "hep-ph/0302104",
    archivePrefix = "arXiv",
    reportNumber = "CERN-TH-2003-026",
    doi = "10.1016/S0370-2693(03)00656-7",
    journal = "Phys. Lett. B",
    volume = "564",
    pages = "65--72",
    year = "2003"
}

@article{Catani:2013tia,
    author = "Catani, Stefano and Cieri, Leandro and de Florian, Daniel and Ferrera, Giancarlo and Grazzini, Massimiliano",
    title = "{Universality of transverse-momentum resummation and hard factors at the NNLO}",
    eprint = "1311.1654",
    archivePrefix = "arXiv",
    primaryClass = "hep-ph",
    reportNumber = "ZU-TH-25-13",
    doi = "10.1016/j.nuclphysb.2014.02.011",
    journal = "Nucl. Phys. B",
    volume = "881",
    pages = "414--443",
    year = "2014"
}

@article{Bozzi:2005wk,
    author = "Bozzi, Giuseppe and Catani, Stefano and de Florian, Daniel and Grazzini, Massimiliano",
    title = "{Transverse-momentum resummation and the spectrum of the Higgs boson at the LHC}",
    eprint = "hep-ph/0508068",
archivePrefix = "arXiv",
    doi = "10.1016/j.nuclphysb.2005.12.022",
    journal = "Nucl. Phys. B",
    volume = "737",
    pages = "73--120",
    year = "2006"
}

@article{Catani:2000vq,
    author = "Catani, Stefano and de Florian, Daniel and Grazzini, Massimiliano",
    title = "{Universality of nonleading logarithmic contributions in transverse momentum distributions}",
    eprint = "hep-ph/0008184",
    archivePrefix = "arXiv",
    reportNumber = "CERN-TH-2000-244",
    doi = "10.1016/S0550-3213(00)00617-9",
    journal = "Nucl. Phys. B",
    volume = "596",
    pages = "299--312",
    year = "2001"
}

@article{Camarda:2019zyx,
    author = "Camarda, Stefano and others",
    title = "{DYTurbo: Fast predictions for Drell-Yan processes}",
    eprint = "1910.07049",
    archivePrefix = "arXiv",
    primaryClass = "hep-ph",
    doi = "10.1140/epjc/s10052-020-7757-5",
    journal = "Eur. Phys. J. C",
    volume = "80",
    number = "3",
    pages = "251",
    year = "2020",
    note = "[Erratum: Eur.Phys.J.C 80, 440 (2020)]"
}

@article{Bizon:2019zgf,
    author = "Bizon, Wojciech and Gehrmann-De Ridder, Aude and Gehrmann, Thomas and Glover, Nigel and Huss, Alexander and Monni, Pier Francesco and Re, Emanuele and Rottoli, Luca and Walker, Duncan M.",
    title = "{The transverse momentum spectrum of weak gauge bosons at N ${}^3$ LL + NNLO}",
    eprint = "1905.05171",
    archivePrefix = "arXiv",
    primaryClass = "hep-ph",
    reportNumber = "ZU-TH 21/1, CERN-TH-2019-050, LAPTH-026/19, IPPP/19/38",
    doi = "10.1140/epjc/s10052-019-7324-0",
    journal = "Eur. Phys. J. C",
    volume = "79",
    number = "10",
    pages = "868",
    year = "2019"
}

@article{Parisi:1979se,
    author = "Parisi, G. and Petronzio, R.",
    title = "{Small Transverse Momentum Distributions in Hard Processes}",
    reportNumber = "CERN-TH-2627",
    doi = "10.1016/0550-3213(79)90040-3",
    journal = "Nucl. Phys. B",
    volume = "154",
    pages = "427--440",
    year = "1979"
}

@article{Bozzi:2010xn,
    author = "Bozzi, Giuseppe and Catani, Stefano and Ferrera, Giancarlo and de Florian, Daniel and Grazzini, Massimiliano",
    title = "{Production of Drell-Yan lepton pairs in hadron collisions: Transverse-momentum resummation at next-to-next-to-leading logarithmic accuracy}",
    eprint = "1007.2351",
    archivePrefix = "arXiv",
    primaryClass = "hep-ph",
    doi = "10.1016/j.physletb.2010.12.024",
    journal = "Phys. Lett. B",
    volume = "696",
    pages = "207--213",
    year = "2011"
}

@article{Gehrmann:2012ze,
    author = "Gehrmann, Thomas and Lubbert, Thomas and Yang, Li Lin",
    title = "{Transverse parton distribution functions at next-to-next-to-leading order: the quark-to-quark case}",
    eprint = "1209.0682",
    archivePrefix = "arXiv",
    primaryClass = "hep-ph",
    doi = "10.1103/PhysRevLett.109.242003",
    journal = "Phys. Rev. Lett.",
    volume = "109",
    pages = "242003",
    year = "2012"
}

@article{NNPDF:2024nan,
    author = "Ball, Richard D. and others",
    collaboration = "NNPDF",
    title = "{The path to $\hbox {N}^3\hbox {LO}$ parton distributions}",
    eprint = "2402.18635",
    archivePrefix = "arXiv",
    primaryClass = "hep-ph",
    reportNumber = "Nikhef-2023-020, TIF-UNIMI-2023-23, Edinburgh 2023/29, CERN-TH-2024-033",
    doi = "10.1140/epjc/s10052-024-12891-7",
    journal = "Eur. Phys. J. C",
    volume = "84",
    number = "7",
    pages = "659",
    year = "2024"
}

@article{McGowan:2022nag,
    author = "McGowan, J. and Cridge, T. and Harland-Lang, L. A. and Thorne, R. S.",
    title = "{Approximate N$^{3}$LO parton distribution functions with theoretical uncertainties: MSHT20aN$^3$LO PDFs}",
    eprint = "2207.04739",
    archivePrefix = "arXiv",
    primaryClass = "hep-ph",
    doi = "10.1140/epjc/s10052-023-11236-0",
    journal = "Eur. Phys. J. C",
    volume = "83",
    number = "3",
    pages = "185",
    year = "2023",
    note = "[Erratum: Eur.Phys.J.C 83, 302 (2023)]"
}

@article{Qiu:2000hf,
    author = "Qiu, Jian-wei and Zhang, Xiao-fei",
    title = "{Role of the nonperturbative input in QCD resummed Drell-Yan $Q_{T}$ distributions}",
    eprint = "hep-ph/0012348",
    archivePrefix = "arXiv",
    doi = "10.1103/PhysRevD.63.114011",
    journal = "Phys. Rev. D",
    volume = "63",
    pages = "114011",
    year = "2001"
}

@article{Catani:1996yz,
    author = "Catani, Stefano and Mangano, Michelangelo L. and Nason, Paolo and Trentadue, Luca",
    title = "{The Resummation of soft gluons in hadronic collisions}",
    eprint = "hep-ph/9604351",
    archivePrefix = "arXiv",
    reportNumber = "CERN-TH-96-86",
    doi = "10.1016/0550-3213(96)00399-9",
    journal = "Nucl. Phys. B",
    volume = "478",
    pages = "273--310",
    year = "1996"
}

@article{Laenen:2000de,
    author = "Laenen, Eric and Sterman, George F. and Vogelsang, Werner",
    title = "{Higher order QCD corrections in prompt photon production}",
    eprint = "hep-ph/0002078",
    archivePrefix = "arXiv",
    reportNumber = "YITP-99-69, NIKHEF-00-002",
    doi = "10.1103/PhysRevLett.84.4296",
    journal = "Phys. Rev. Lett.",
    volume = "84",
    pages = "4296--4299",
    year = "2000"
}

@article{Lustermans:2019plv,
    author = "Lustermans, Gillian and Michel, Johannes K. L. and Tackmann, Frank J. and Waalewijn, Wouter J.",
    title = "{Joint two-dimensional resummation in $q_{T}$ and $0$-jettiness at NNLL}",
    eprint = "1901.03331",
    archivePrefix = "arXiv",
    primaryClass = "hep-ph",
    reportNumber = "DESY-19-004, DESY 19-004, NIKHEF 2018-053",
    doi = "10.1007/JHEP03(2019)124",
    journal = "JHEP",
    volume = "03",
    pages = "124",
    year = "2019"
}

@article{Collins:1984kg,
    author = "Collins, John C. and Soper, Davison E. and Sterman, George F.",
    title = "{Transverse Momentum Distribution in Drell-Yan Pair and W and Z Boson Production}",
    reportNumber = "CERN-TH-3923",
    doi = "10.1016/0550-3213(85)90479-1",
    journal = "Nucl. Phys. B",
    volume = "250",
    pages = "199--224",
    year = "1985"
}

@article{Collins:2014jpa,
    author = "Collins, John and Rogers, Ted",
    title = "{Understanding the large-distance behavior of transverse-momentum-dependent parton densities and the Collins-Soper evolution kernel}",
    eprint = "1412.3820",
    archivePrefix = "arXiv",
    primaryClass = "hep-ph",
    reportNumber = "YITP-WSB-14-45, JLAB-THY-15-2001",
    doi = "10.1103/PhysRevD.91.074020",
    journal = "Phys. Rev. D",
    volume = "91",
    number = "7",
    pages = "074020",
    year = "2015"
}

@article{Dokshitzer:1978yd,
    author = "Dokshitzer, Yuri L. and Diakonov, Dmitri and Troian, S. I.",
    title = "{On the Transverse Momentum Distribution of Massive Lepton Pairs}",
    doi = "10.1016/0370-2693(78)90240-X",
    journal = "Phys. Lett. B",
    volume = "79",
    pages = "269--272",
    year = "1978"
}

@article{Collins:1989gx,
    author = "Collins, John C. and Soper, Davison E. and Sterman, George F.",
    title = "{Factorization of Hard Processes in QCD}",
    eprint = "hep-ph/0409313",
    archivePrefix = "arXiv",
    reportNumber = "ITP-SB-89-31",
    doi = "10.1142/9789814503266_0001",
    journal = "Adv. Ser. Direct. High Energy Phys.",
    volume = "5",
    pages = "1--91",
    year = "1989"
}

@article{Collins:1981uk,
    author = "Collins, John C. and Soper, Davison E.",
    title = "{Back-To-Back Jets in QCD}",
    reportNumber = "OITS-155",
    doi = "10.1016/0550-3213(81)90339-4",
    journal = "Nucl. Phys. B",
    volume = "193",
    pages = "381",
    year = "1981",
    note = "[Erratum: Nucl.Phys.B 213, 545 (1983)]"
}

@article{Collins:1985ue,
    author = "Collins, John C. and Soper, Davison E. and Sterman, George F.",
    title = "{Factorization for Short Distance Hadron - Hadron Scattering}",
    reportNumber = "OITS-287",
    doi = "10.1016/0550-3213(85)90565-6",
    journal = "Nucl. Phys. B",
    volume = "261",
    pages = "104--142",
    year = "1985"
}

@article{Bacchetta:2025ara,
    author = "Bacchetta, Alessandro and Bertone, Valerio and Bissolotti, Chiara and Cerutti, Matteo and Radici, Marco and Rodini, Simone and Rossi, Lorenzo",
    collaboration = "MAP (Multi-dimensional Analyses of Partonic distributions)",
    title = "{Neural-Network Extraction of Unpolarized Transverse-Momentum-Dependent Distributions}",
    eprint = "2502.04166",
    archivePrefix = "arXiv",
    primaryClass = "hep-ph",
    reportNumber = "DESY-25-022, JLAB-THY-25-4221",
    doi = "10.1103/csc2-bj91",
    journal = "Phys. Rev. Lett.",
    volume = "135",
    number = "2",
    pages = "021904",
    year = "2025"
}

@article{Bacchetta:2024qre,
    author = "Bacchetta, Alessandro and Bertone, Valerio and Bissolotti, Chiara and Bozzi, Giuseppe and Cerutti, Matteo and Delcarro, Filippo and Radici, Marco and Rossi, Lorenzo and Signori, Andrea",
    collaboration = "MAP (Multi-dimensional Analyses of Partonic distributions)",
    title = "{Flavor dependence of unpolarized quark transverse momentum distributions from a global fit}",
    eprint = "2405.13833",
    archivePrefix = "arXiv",
    primaryClass = "hep-ph",
    reportNumber = "JLAB-THY-24-4066",
    doi = "10.1007/JHEP08(2024)232",
    journal = "JHEP",
    volume = "08",
    pages = "232",
    year = "2024"
}

@article{Moos:2025sal,
    author = "Moos, Valentin and Scimemi, Ignazio and Vladimirov, Alexey and Zurita, Pia",
    title = "{Determination of unpolarized TMD distributions from the fit of Drell-Yan and SIDIS data at N$^4$LL}",
    eprint = "2503.11201",
    archivePrefix = "arXiv",
    primaryClass = "hep-ph",
    reportNumber = "IPARCOS-UCM-25-018",
    month = "3",
    year = "2025"
}

@article{Neumann:2022lft,
    author = "Neumann, Tobias and Campbell, John",
    title = "{Fiducial Drell-Yan production at the LHC improved by transverse-momentum resummation at N4LLp+N3LO}",
    eprint = "2207.07056",
    archivePrefix = "arXiv",
    primaryClass = "hep-ph",
    reportNumber = "FERMILAB-PUB-22-528-T",
    doi = "10.1103/PhysRevD.107.L011506",
    journal = "Phys. Rev. D",
    volume = "107",
    number = "1",
    pages = "L011506",
    year = "2023"
}

@article{Baglio:2022wzu,
    author = "Baglio, Julien and Duhr, Claude and Mistlberger, Bernhard and Szafron, Robert",
    title = "{Inclusive production cross sections at N$^{3}$LO}",
    eprint = "2209.06138",
    archivePrefix = "arXiv",
    primaryClass = "hep-ph",
    reportNumber = "CERN-TH-2022-109, SLAC-PUB-17699, BONN-TH-2022-22",
    doi = "10.1007/JHEP12(2022)066",
    journal = "JHEP",
    volume = "12",
    pages = "066",
    year = "2022"
}

@article{Kulesza:2002rh,
    author = "Kulesza, Anna and Sterman, George F. and Vogelsang, Werner",
    title = "{Joint resummation in electroweak boson production}",
    eprint = "hep-ph/0202251",
    archivePrefix = "arXiv",
    reportNumber = "BNL-HET-01-43, BNL-NT-01-32, RBRC-232, YITP-SB-01-75",
    doi = "10.1103/PhysRevD.66.014011",
    journal = "Phys. Rev. D",
    volume = "66",
    pages = "014011",
    year = "2002"
}

@article{Kulesza:2003wn,
    author = "Kulesza, Anna and Sterman, George F. and Vogelsang, Werner",
    title = "{Joint resummation for Higgs production}",
    eprint = "hep-ph/0309264",
    archivePrefix = "arXiv",
    reportNumber = "BNL-HET-03-20, BNL-NT-03-26, RBRC-335, YITP-SB-03-47",
    doi = "10.1103/PhysRevD.69.014012",
    journal = "Phys. Rev. D",
    volume = "69",
    pages = "014012",
    year = "2004"
}

@article{Moreno:1990sf,
    author = "Moreno, G. and others",
    title = "{Dimuon Production in Proton - Copper Collisions at $\sqrt{s}$ = 38.8-GeV}",
    reportNumber = "FERMILAB-PUB-90-223-E",
    doi = "10.1103/PhysRevD.43.2815",
    journal = "Phys. Rev. D",
    volume = "43",
    pages = "2815--2836",
    year = "1991"
}

@article{Ito:1980ev,
    author = "Ito, A. S. and others",
    title = "{Measurement of the Continuum of Dimuons Produced in High-Energy Proton - Nucleus Collisions}",
    reportNumber = "FERMILAB-PUB-80-019-E",
    doi = "10.1103/PhysRevD.23.604",
    journal = "Phys. Rev. D",
    volume = "23",
    pages = "604--633",
    year = "1981"
}

@article{CDF:2012brb,
    author = "Aaltonen, T. and others",
    collaboration = "CDF",
    title = "{Transverse momentum cross section of $e^+e^-$ pairs in the $Z$-boson region from $p\bar{p}$ collisions at $\sqrt{s}=1.96$ TeV}",
    eprint = "1207.7138",
    archivePrefix = "arXiv",
    primaryClass = "hep-ex",
    reportNumber = "FERMILAB-PUB-12-421-E",
    doi = "10.1103/PhysRevD.86.052010",
    journal = "Phys. Rev. D",
    volume = "86",
    pages = "052010",
    year = "2012"
}

@article{ATLAS:2023lsr,
    author = "Aad, Georges and others",
    collaboration = "ATLAS",
    title = "{A precise measurement of the Z-boson double-differential transverse momentum and rapidity distributions in the full phase space of the decay leptons with the ATLAS experiment at~$\sqrt{s}=8$~TeV}",
    eprint = "2309.09318",
    archivePrefix = "arXiv",
    primaryClass = "hep-ex",
    reportNumber = "CERN-EP-2023-171",
    doi = "10.1140/epjc/s10052-024-12438-w",
    journal = "Eur. Phys. J. C",
    volume = "84",
    number = "3",
    pages = "315",
    year = "2024"
}

@article{Alekhin:2014irh,
    author = "Alekhin, S. and others",
    title = "{HERAFitter}",
    eprint = "1410.4412",
    archivePrefix = "arXiv",
    primaryClass = "hep-ph",
    reportNumber = "DESY-14-188, DESY-REPORT-14-188, FERMILAB-PUB-14-603-CMS",
    doi = "10.1140/epjc/s10052-015-3480-z",
    journal = "Eur. Phys. J. C",
    volume = "75",
    number = "7",
    pages = "304",
    year = "2015"
}

@article{LatticePartonLPC:2023pdv,
    author = "Chu, Min-Huan and others",
    collaboration = "Lattice Parton (LPC)",
    title = "{Lattice calculation of the intrinsic soft function and the Collins-Soper kernel}",
    eprint = "2306.06488",
    archivePrefix = "arXiv",
    primaryClass = "hep-lat",
    doi = "10.1007/JHEP08(2023)172",
    journal = "JHEP",
    volume = "08",
    pages = "172",
    year = "2023"
}

@article{Avkhadiev:2023poz,
    author = "Avkhadiev, Artur and Shanahan, Phiala E. and Wagman, Michael L. and Zhao, Yong",
    title = "{Collins-Soper kernel from lattice QCD at the physical pion mass}",
    eprint = "2307.12359",
    archivePrefix = "arXiv",
    primaryClass = "hep-lat",
    reportNumber = "MIT-CTP/5587, FERMILAB-PUB-23-375-T",
    doi = "10.1103/PhysRevD.108.114505",
    journal = "Phys. Rev. D",
    volume = "108",
    number = "11",
    pages = "114505",
    year = "2023"
}

@article{Avkhadiev:2024mgd,
    author = "Avkhadiev, Artur and Shanahan, Phiala E. and Wagman, Michael L. and Zhao, Yong",
    title = "{Determination of the Collins-Soper Kernel from Lattice QCD}",
    eprint = "2402.06725",
    archivePrefix = "arXiv",
    primaryClass = "hep-lat",
    reportNumber = "FERMILAB-PUB-24-0037-T, MIT-CTP/5677",
    doi = "10.1103/PhysRevLett.132.231901",
    journal = "Phys. Rev. Lett.",
    volume = "132",
    number = "23",
    pages = "231901",
    year = "2024"
}

@article{Bollweg:2024zet,
    author = "Bollweg, Dennis and Gao, Xiang and Mukherjee, Swagato and Zhao, Yong",
    title = "{Nonperturbative Collins-Soper kernel from chiral quarks with physical masses}",
    eprint = "2403.00664",
    archivePrefix = "arXiv",
    primaryClass = "hep-lat",
    doi = "10.1016/j.physletb.2024.138617",
    journal = "Phys. Lett. B",
    volume = "852",
    pages = "138617",
    year = "2024"
}

@article{Bollweg:2025iol,
    author = "Bollweg, Dennis and Gao, Xiang and He, Jinchen and Mukherjee, Swagato and Zhao, Yong",
    title = "{Transverse-momentum-dependent pion structures from lattice QCD: Collins-Soper kernel, soft factor, TMDWF, and TMDPDF}",
    eprint = "2504.04625",
    archivePrefix = "arXiv",
    primaryClass = "hep-lat",
    doi = "10.1103/j3n6-8kxy",
    journal = "Phys. Rev. D",
    volume = "112",
    number = "3",
    pages = "034501",
    year = "2025"
}

@misc{WEBLINK,
  author = "Camarda, Stefano and others",
  howpublished = {\url{https://dyturbo.hepforge.org}}
}

@article{Moos:2023yfa,
    author = "Moos, Valentin and Scimemi, Ignazio and Vladimirov, Alexey and Zurita, Pia",
    title = "{Extraction of unpolarized transverse momentum distributions from the fit of Drell-Yan data at N$^{4}$LL}",
    eprint = "2305.07473",
    archivePrefix = "arXiv",
    primaryClass = "hep-ph",
    reportNumber = "IPARCOS-UCM-035",
    doi = "10.1007/JHEP05(2024)036",
    journal = "JHEP",
    volume = "05",
    pages = "036",
    year = "2024"
}

@book{Collins:2011zzd,
    author = "Collins, John",
    title = "{Foundations of Perturbative QCD}",
    doi = "10.1017/9781009401845",
    isbn = "978-1-009-40184-5, 978-1-009-40183-8, 978-1-009-40182-1",
    publisher = "Cambridge University Press",
    volume = "32",
    year = "2011"
}

@article{Echevarria:2011epo,
    author = "Echevarria, Miguel G. and Idilbi, Ahmad and Scimemi, Ignazio",
    title = "{Factorization Theorem For Drell-Yan At Low $q_T$ And Transverse Momentum Distributions On-The-Light-Cone}",
    eprint = "1111.4996",
    archivePrefix = "arXiv",
    primaryClass = "hep-ph",
    doi = "10.1007/JHEP07(2012)002",
    journal = "JHEP",
    volume = "07",
    pages = "002",
    year = "2012"
}

@article{Catani:2018krb,
    author = "Catani, Stefano and Cieri, Leandro and de Florian, Daniel and Ferrera, Giancarlo and Grazzini, Massimiliano",
    title = "{Diphoton production at the LHC: a QCD study up to NNLO}",
    eprint = "1802.02095",
    archivePrefix = "arXiv",
    primaryClass = "hep-ph",
    reportNumber = "ZH-TH-06-18",
    doi = "10.1007/JHEP04(2018)142",
    journal = "JHEP",
    volume = "04",
    pages = "142",
    year = "2018"
}

@article{CMS:2011utm,
    author = "Chatrchyan, Serguei and others",
    collaboration = "CMS",
    title = "{Measurement of the weak mixing angle with the Drell-Yan process in proton-proton collisions at the LHC}",
    eprint = "1110.2682",
    archivePrefix = "arXiv",
    primaryClass = "hep-ex",
    reportNumber = "CMS-EWK-11-003, CERN-PH-EP-2011-159",
    doi = "10.1103/PhysRevD.84.112002",
    journal = "Phys. Rev. D",
    volume = "84",
    pages = "112002",
    year = "2011"
}

@article{Camarda:2016twt,
    author = "Camarda, Stefano and Cuth, Jakub and Schott, Matthias",
    title = "{Determination of the muonic branching ratio of the W boson and its total width via cross-section measurements at the Tevatron and LHC}",
    eprint = "1607.05084",
    archivePrefix = "arXiv",
    primaryClass = "hep-ex",
    doi = "10.1140/epjc/s10052-016-4461-6",
    journal = "Eur. Phys. J. C",
    volume = "76",
    number = "11",
    pages = "613",
    year = "2016"
}

@article{CDF:2022hxs,
    author = "Aaltonen, T. and others",
    collaboration = "CDF",
    title = "{High-precision measurement of the $W$          boson mass with the CDF II detector}",
    reportNumber = "FERMILAB-PUB-22-254-PPD",
    doi = "10.1126/science.abk1781",
    journal = "Science",
    volume = "376",
    number = "6589",
    pages = "170--176",
    year = "2022"
}

@article{ATLAS:2017rzl,
    author = "Aaboud, Morad and others",
    collaboration = "ATLAS",
    title = "{Measurement of the $W$-boson mass in pp collisions at $\sqrt{s}=7$ TeV with the ATLAS detector}",
    eprint = "1701.07240",
    archivePrefix = "arXiv",
    primaryClass = "hep-ex",
    reportNumber = "CERN-EP-2016-305",
    doi = "10.1140/epjc/s10052-017-5475-4",
    journal = "Eur. Phys. J. C",
    volume = "78",
    number = "2",
    pages = "110",
    year = "2018",
    note = "[Erratum: Eur.Phys.J.C 78, 898 (2018)]"
}

@article{Rottoli:2023xdc,
    author = "Rottoli, Luca and Torrielli, Paolo and Vicini, Alessandro",
    title = "{Determination of the W-boson mass at hadron colliders}",
    eprint = "2301.04059",
    archivePrefix = "arXiv",
    primaryClass = "hep-ph",
    reportNumber = "ZU-TH 01/23, TIF-UNIMI-2023-1",
    doi = "10.1140/epjc/s10052-023-12128-z",
    journal = "Eur. Phys. J. C",
    volume = "83",
    number = "10",
    pages = "948",
    year = "2023"
}

@article{CDF:2013dpa,
    author = "Aaltonen, Timo Antero and others",
    collaboration = "CDF, D0",
    title = "{Combination of CDF and D0 $W$-Boson Mass Measurements}",
    eprint = "1307.7627",
    archivePrefix = "arXiv",
    primaryClass = "hep-ex",
    reportNumber = "FERMILAB-PUB-13-289-E",
    doi = "10.1103/PhysRevD.88.052018",
    journal = "Phys. Rev. D",
    volume = "88",
    number = "5",
    pages = "052018",
    year = "2013"
}

@article{LHCb:2021bjt,
    author = "Aaij, Roel and others",
    collaboration = "LHCb",
    title = "{Measurement of the W boson mass}",
    eprint = "2109.01113",
    archivePrefix = "arXiv",
    primaryClass = "hep-ex",
    reportNumber = "LHCb-PAPER-2021-024, CERN-EP-2021-170",
    doi = "10.1007/JHEP01(2022)036",
    journal = "JHEP",
    volume = "01",
    pages = "036",
    year = "2022"
}

@article{Holmes:2011ey,
    author = "Holmes, Stephen and Moore, Ronald S. and Shiltsev, Vladimir",
    title = "{Overview of the Tevatron Collider Complex: Goals, Operations and Performance}",
    eprint = "1106.0909",
    archivePrefix = "arXiv",
    primaryClass = "physics.acc-ph",
    reportNumber = "FERMILAB-PUB-11-245-AD-APC",
    doi = "10.1088/1748-0221/6/08/T08001",
    journal = "JINST",
    volume = "6",
    pages = "T08001",
    year = "2011"
}

@article{Evans:2008zzb,
    editor = "Evans, Lyndon and Bryant, Philip",
    title = "{LHC Machine}",
    doi = "10.1088/1748-0221/3/08/S08001",
    journal = "JINST",
    volume = "3",
    pages = "S08001",
    year = "2008"
}

@article{CidVidal:2018eel,
    author = "Cid Vidal, Xabier and others",
    editor = "Dainese, Andrea and Mangano, Michelangelo and Meyer, Andreas B. and Nisati, Aleandro and Salam, Gavin and Vesterinen, Mika Anton",
    title = "{Report from Working Group 3}: {Beyond the Standard Model physics at the HL-LHC and HE-LHC}",
    eprint = "1812.07831",
    archivePrefix = "arXiv",
    primaryClass = "hep-ph",
    reportNumber = "CERN-LPCC-2018-05",
    doi = "10.23731/CYRM-2019-007.585",
    journal = "CERN Yellow Rep. Monogr.",
    volume = "7",
    pages = "585--865",
    year = "2019"
}

@article{Bozzi:2007pn,
    author = "Bozzi, Giuseppe and Catani, Stefano and de Florian, Daniel and Grazzini, Massimiliano",
    title = "{Higgs boson production at the LHC: Transverse-momentum resummation and rapidity dependence}",
    eprint = "0705.3887",
    archivePrefix = "arXiv",
    primaryClass = "hep-ph",
    reportNumber = "SFB-CPP-07-18, KA-TP-13-2007",
    doi = "10.1016/j.nuclphysb.2007.09.034",
    journal = "Nucl. Phys. B",
    volume = "791",
    pages = "1--19",
    year = "2008"
}

@article{Cieri:2015rqa,
    author = "Cieri, Leandro and Coradeschi, Francesco and de Florian, Daniel",
    title = "{Diphoton production at hadron colliders: transverse-momentum resummation at next-to-next-to-leading logarithmic accuracy}",
    eprint = "1505.03162",
    archivePrefix = "arXiv",
    primaryClass = "hep-ph",
    doi = "10.1007/JHEP06(2015)185",
    journal = "JHEP",
    volume = "06",
    pages = "185",
    year = "2015"
}

@article{Grazzini:2015wpa,
    author = "Grazzini, Massimiliano and Kallweit, Stefan and Rathlev, Dirk and Wiesemann, Marius",
    title = "{Transverse-momentum resummation for vector-boson pair production at NNLL+NNLO}",
    eprint = "1507.02565",
    archivePrefix = "arXiv",
    primaryClass = "hep-ph",
    reportNumber = "ZU-TH-20-15, MITP-15-051",
    doi = "10.1007/JHEP08(2015)154",
    journal = "JHEP",
    volume = "08",
    pages = "154",
    year = "2015"
}

@article{deFlorian:2011xf,
    author = "de Florian, Daniel and Ferrera, Giancarlo and Grazzini, Massimiliano and Tommasini, Damiano",
    title = "{Transverse-momentum resummation: Higgs boson production at the Tevatron and the LHC}",
    eprint = "1109.2109",
    archivePrefix = "arXiv",
    primaryClass = "hep-ph",
    reportNumber = "ZU-TH-17-11",
    doi = "10.1007/JHEP11(2011)064",
    journal = "JHEP",
    volume = "11",
    pages = "064",
    year = "2011"
}

@article{deFlorian:2012mx,
    author = "de Florian, D. and Ferrera, G. and Grazzini, M. and Tommasini, D.",
    title = "{Higgs boson production at the LHC: transverse momentum resummation effects in the H-{\ensuremath{>}}2gamma, H-{\ensuremath{>}}WW-{\ensuremath{>}}lnu lnu and H-{\ensuremath{>}}ZZ-{\ensuremath{>}}4l decay modes}",
    eprint = "1203.6321",
    archivePrefix = "arXiv",
    primaryClass = "hep-ph",
    reportNumber = "ZU-TH-04-12, IFUM-993-FT",
    doi = "10.1007/JHEP06(2012)132",
    journal = "JHEP",
    volume = "06",
    pages = "132",
    year = "2012"
}

@article{Bacchetta:2019tcu,
    author = "Bacchetta, Alessandro and Bozzi, Giuseppe and Lambertsen, Martin and Piacenza, Fulvio and Steiglechner, Julius and Vogelsang, Werner",
    title = "{Difficulties in the description of Drell-Yan processes at moderate invariant mass and high transverse momentum}",
    eprint = "1901.06916",
    archivePrefix = "arXiv",
    primaryClass = "hep-ph",
    reportNumber = "INT-PUB-19-002",
    doi = "10.1103/PhysRevD.100.014018",
    journal = "Phys. Rev. D",
    volume = "100",
    number = "1",
    pages = "014018",
    year = "2019"
}

@article{Gauld:2021pkr,
    author = "Gauld, R. and Gehrmann-De Ridder, A. and Gehrmann, T. and Glover, E. W. N. and Huss, A. and Majer, I. and Rodriguez Garcia, A.",
    title = "{Transverse momentum distributions in low-mass Drell-Yan lepton pair production at NNLO QCD}",
    eprint = "2110.15839",
    archivePrefix = "arXiv",
    primaryClass = "hep-ph",
    reportNumber = "NIKHEF 2021-027; BONN-TH-2021-10; ZU-TH 51/21; IPPP/21/40;
  CERN-TH-2021-158, NIKHEF 2021-027; BONN-TH-2021-10; ZU-TH 51/21; IPPP/21/40;
CERN-TH-2021-158",
    doi = "10.1016/j.physletb.2022.137111",
    journal = "Phys. Lett. B",
    volume = "829",
    pages = "137111",
    year = "2022"
}

@article{Camarda:2023dqn,
    author = "Camarda, Stefano and Cieri, Leandro and Ferrera, Giancarlo",
    title = "{Drell{\textendash}Yan lepton-pair production: qT resummation at N4LL accuracy}",
    eprint = "2303.12781",
    archivePrefix = "arXiv",
    primaryClass = "hep-ph",
    doi = "10.1016/j.physletb.2023.138125",
    journal = "Phys. Lett. B",
    volume = "845",
    pages = "138125",
    year = "2023"
}

@article{Ebert:2016gcn,
    author = "Ebert, Markus A. and Tackmann, Frank J.",
    title = "{Resummation of Transverse Momentum Distributions in Distribution Space}",
    eprint = "1611.08610",
    archivePrefix = "arXiv",
    primaryClass = "hep-ph",
    reportNumber = "DESY-16-215",
    doi = "10.1007/JHEP02(2017)110",
    journal = "JHEP",
    volume = "02",
    pages = "110",
    year = "2017"
}

@article{Catani:2010pd,
    author = "Catani, Stefano and Grazzini, Massimiliano",
    title = "{QCD transverse-momentum resummation in gluon fusion processes}",
    eprint = "1011.3918",
    archivePrefix = "arXiv",
    primaryClass = "hep-ph",
    doi = "10.1016/j.nuclphysb.2010.12.007",
    journal = "Nucl. Phys. B",
    volume = "845",
    pages = "297--323",
    year = "2011"
}

@article{Becher:2010tm,
    author = "Becher, Thomas and Neubert, Matthias",
    title = "{Drell-Yan Production at Small $q_T$, Transverse Parton Distributions and the Collinear Anomaly}",
    eprint = "1007.4005",
    archivePrefix = "arXiv",
    primaryClass = "hep-ph",
    reportNumber = "HD-THEP-10-13, MZ-TH-10-26",
    doi = "10.1140/epjc/s10052-011-1665-7",
    journal = "Eur. Phys. J. C",
    volume = "71",
    pages = "1665",
    year = "2011"
}

@article{Scimemi:2017etj,
    author = "Scimemi, Ignazio and Vladimirov, Alexey",
    title = "{Analysis of vector boson production within TMD factorization}",
    eprint = "1706.01473",
    archivePrefix = "arXiv",
    primaryClass = "hep-ph",
    doi = "10.1140/epjc/s10052-018-5557-y",
    journal = "Eur. Phys. J. C",
    volume = "78",
    number = "2",
    pages = "89",
    year = "2018"
}

@article{Bizon:2018foh,
    author = "Bizo{\'n}, Wojciech and Chen, Xuan and Gehrmann-De Ridder, Aude and Gehrmann, Thomas and Glover, Nigel and Huss, Alexander and Monni, Pier Francesco and Re, Emanuele and Rottoli, Luca and Torrielli, Paolo",
    title = "{Fiducial distributions in Higgs and Drell-Yan production at N$^{3}$LL+NNLO}",
    eprint = "1805.05916",
    archivePrefix = "arXiv",
    primaryClass = "hep-ph",
    reportNumber = "CERN-TH-2018-105, IPPP/18/34, LAPTH-015/18, OUTP-17-19P, ZU-TH 17/18, IPPP-18-34, LAPTH-015-18, ZU-TH-17-18",
    doi = "10.1007/JHEP12(2018)132",
    journal = "JHEP",
    volume = "12",
    pages = "132",
    year = "2018"
}

@article{Becher:2019bnm,
    author = "Becher, Thomas and Hager, Monika",
    title = "{Event-Based Transverse Momentum Resummation}",
    eprint = "1904.08325",
    archivePrefix = "arXiv",
    primaryClass = "hep-ph",
    doi = "10.1140/epjc/s10052-019-7136-2",
    journal = "Eur. Phys. J. C",
    volume = "79",
    number = "8",
    pages = "665",
    year = "2019"
}

@article{Alioli:2021qbf,
    author = "Alioli, Simone and Bauer, Christian W. and Broggio, Alessandro and Gavardi, Alessandro and Kallweit, Stefan and Lim, Matthew A. and Nagar, Riccardo and Napoletano, Davide and Rottoli, Luca",
    title = "{Matching NNLO predictions to parton showers using N3LL color-singlet transverse momentum resummation in geneva}",
    eprint = "2102.08390",
    archivePrefix = "arXiv",
    primaryClass = "hep-ph",
    reportNumber = "ZU-TH 8/21",
    doi = "10.1103/PhysRevD.104.094020",
    journal = "Phys. Rev. D",
    volume = "104",
    number = "9",
    pages = "094020",
    year = "2021"
}

@article{LatticePartonLPC:2022eev,
    author = "Chu, Min-Huan and others",
    collaboration = "Lattice Parton (LPC)",
    title = "{Nonperturbative determination of the Collins-Soper kernel from quasitransverse-momentum-dependent wave functions}",
    eprint = "2204.00200",
    archivePrefix = "arXiv",
    primaryClass = "hep-lat",
    doi = "10.1103/PhysRevD.106.034509",
    journal = "Phys. Rev. D",
    volume = "106",
    number = "3",
    pages = "034509",
    year = "2022"
}

@article{Billis:2024dqq,
    author = "Billis, Georgios and Michel, Johannes K. L. and Tackmann, Frank J.",
    title = "{Drell-Yan transverse-momentum spectra at N$^{3}$LL$^{'}$ and approximate N$^{4}$LL with SCETlib}",
    eprint = "2411.16004",
    archivePrefix = "arXiv",
    primaryClass = "hep-ph",
    reportNumber = "DESY-23-081, MIT-CTP 5572, Nikhef 2024-007",
    doi = "10.1007/JHEP02(2025)170",
    journal = "JHEP",
    volume = "02",
    pages = "170",
    year = "2025"
}

@article{Bacchetta:2022awv,
    author = "Bacchetta, Alessandro and Bertone, Valerio and Bissolotti, Chiara and Bozzi, Giuseppe and Cerutti, Matteo and Piacenza, Fulvio and Radici, Marco and Signori, Andrea",
    collaboration = "MAP (Multi-dimensional Analyses of Partonic distributions)",
    title = "{Unpolarized transverse momentum distributions from a global fit of Drell-Yan and semi-inclusive deep-inelastic scattering data}",
    eprint = "2206.07598",
    archivePrefix = "arXiv",
    primaryClass = "hep-ph",
    doi = "10.1007/JHEP10(2022)127",
    journal = "JHEP",
    volume = "10",
    pages = "127",
    year = "2022"
}

@article{Aslan:2024nqg,
    author = "Aslan, F. and Boglione, M. and Gonzalez-Hernandez, J. O. and Rainaldi, T. and Rogers, T. C. and Simonelli, A.",
    title = "{Phenomenology of TMD parton distributions in Drell-Yan and Z0 boson production in a hadron structure oriented approach}",
    eprint = "2401.14266",
    archivePrefix = "arXiv",
    primaryClass = "hep-ph",
    reportNumber = "JLAB-THY-24-3987",
    doi = "10.1103/PhysRevD.110.074016",
    journal = "Phys. Rev. D",
    volume = "110",
    number = "7",
    pages = "074016",
    year = "2024"
}

@article{Cuerpo:2025zde,
    author = "Cuerpo, Alejandro Bris and Scimemi, Ignazio and Vladimirov, Alexey",
    title = "{Assessing the sensitivity of Energy-Energy Correlations in $e^+e^-$ annihilation to TMD dynamics}",
    eprint = "2507.17478",
    archivePrefix = "arXiv",
    primaryClass = "hep-ph",
    reportNumber = "IPARCOS-UCM-25-028",
    month = "7",
    year = "2025"
}

@article{ATLAS:2023lhg,
    author = "Aad, Georges and others",
    collaboration = "ATLAS",
    title = "{A precise determination of the strong-coupling constant from the recoil of $Z$ bosons with the ATLAS experiment at $\sqrt{s} = 8$ TeV}",
    eprint = "2309.12986",
    archivePrefix = "arXiv",
    primaryClass = "hep-ex",
    month = "9",
    year = "2023"
}

@article{Boughezal:2015ded,
    author = "Boughezal, Radja and Campbell, John M. and Ellis, R. Keith and Focke, Christfried and Giele, Walter T. and Liu, Xiaohui and Petriello, Frank",
    title = "{Z-boson production in association with a jet at next-to-next-to-leading order in perturbative QCD}",
    eprint = "1512.01291",
    archivePrefix = "arXiv",
    primaryClass = "hep-ph",
    reportNumber = "FERMILAB-PUB-15-519-T, IPPP-15-79",
    doi = "10.1103/PhysRevLett.116.152001",
    journal = "Phys. Rev. Lett.",
    volume = "116",
    number = "15",
    pages = "152001",
    year = "2016"
}

@article{Catani:2022sgr,
    author = "Catani, Stefano and Dhani, Prasanna K.",
    title = "{Collinear functions for QCD resummations}",
    eprint = "2208.05840",
    archivePrefix = "arXiv",
    primaryClass = "hep-ph",
    doi = "10.1007/JHEP03(2023)200",
    journal = "JHEP",
    volume = "03",
    pages = "200",
    year = "2023"
}

\end{document}